\documentclass[longbibliography,showpacs,preprintnumbers,amsmath,amssymb,aps,prd,nofootinbib,superscriptaddress,12pt]{revtex4-2}
\usepackage{graphicx}
\usepackage{caption}
\usepackage{subcaption}
\usepackage{xcolor}
\usepackage{hyperref}
\usepackage{amsfonts}
\usepackage{bm}
\usepackage[a4paper, margin=2cm]{geometry}
\usepackage{mathrsfs}
\usepackage{soul}
\usepackage{makecell,multirow}
\usepackage{rotating}
\usepackage[utf8]{inputenc}
\usepackage{amsmath}
\usepackage{makecell}

\usepackage{amssymb}
\usepackage{tensor}
\usepackage{graphicx}
\usepackage{bigints}
\usepackage{bbm}
\usepackage{graphicx}
\usepackage{subcaption}
\usepackage{epsf}
\usepackage{bm}
\usepackage{amsmath}
\usepackage{amsfonts}
\usepackage{amssymb}
\usepackage{graphicx}
\usepackage{tabularx}
\usepackage{multirow}
\usepackage{color}%
\providecommand{\U}[1]{\protect\rule{.1in}{.1in}}

\newcommand{\ie}{\begin{equation}}
\newcommand{\fe}{\end{equation}}

\newcommand{\mincir}{\raise
-3.truept\hbox{\rlap{\hbox{$\sim$}}\raise4.truept\hbox{$<$}\ }}
\newcommand{\magcir}{\raise
-3.truept\hbox{\rlap{\hbox{$\sim$}}\raise4.truept\hbox{$>$}\ }}

\providecommand{\U}[1]{\protect\rule{.1in}{.1in}}

\usepackage{tikz,xcolor,hyperref}

\usepackage{hyperref}             
\hypersetup{
    colorlinks=true,              
    breaklinks=true,              
    citecolor=blue,               
    linkcolor=[rgb]{0,0.5,0.9},   
    urlcolor=red,                 
    filecolor=green               
}

\definecolor{lime}{HTML}{A6CE39}
\DeclareRobustCommand{\orcidicon}{%
	\begin{tikzpicture}
	\draw[lime, fill=lime] (0,0) 
	circle [radius=0.16] 
	node[white] {{\fontfamily{qag}\selectfont \tiny ID}};
	\draw[white, fill=white] (-0.0625,0.095) 
	circle [radius=0.007];
	\end{tikzpicture}
	\hspace{-2mm}
}

\foreach \x in {A, ..., Z}{%
	\expandafter\xdef\csname orcid\x\endcsname{\noexpand\href{https://orcid.org/\csname orcidauthor\x\endcsname}{\noexpand\orcidicon}}
}

\begin{document}

\title{\Large{Light propagation and quasinormal modes of a topologically charged Schwarzschild–Klinkhamer wormhole}}

\author{C. F. S. Pereira}
	\email{carlosfisica32@gmail.com}
	\affiliation{Departamento de F\'isica e Qu\'imica, Universidade Federal do Esp\'irito Santo, Av.Fernando Ferrari, 514, Goiabeiras, Vit\'oria, ES 29060-900, Brazil.}

\author{H. Belich} 
\email{humberto.belich@ufes.br}
\affiliation{Departamento de F\'isica e Qu\'imica, Universidade Federal do Esp\'irito Santo, Av.Fernando Ferrari, 514, Goiabeiras, Vit\'oria, ES 29060-900, Brazil.}        
	\author{A. R. Soares}
    \email{adriano.soares@ifma.edu.br}
	\affiliation{Grupo de Estudos e Pesquisas em Laborat\'orio de Educa\c{c}\~ao matem\'atica, Instituto Federal de Educa\c{c}\~ao Ci\^encia e Tecnologia do Maranh\~ao,  R. Dep. Gast\~ao Vieira, 1000, CEP 65393-000 Buriticupu, MA, Brazil.}

    \author{Marcos V. de S. Silva}
	\email{marcos.sousa@uva.es}
\affiliation{Department of Theoretical Physics, Atomic and Optics, Campus Miguel Delibes, \\ University of Valladolid UVA, Paseo Bel\'en, 7,
47011 - Valladolid, Spain}
\author{R. L. L. Vit\'oria}
    \email{ricardovitoria@professor.uema.br/ricardo-luis91@hotmail.com}
	\affiliation{Programa de Pós-Gradua\c c\~ao em Engenharia Aeroespacial, Universidade Estadual do Maranh\~ao, Cidade Universit\'aria Paulo VI, São Lu\'is 65055-310, MA, Brazil}
    \affiliation{Faculdade de F\'isica, Universidade Federal do Par\'a, Av. Augusto Corr\^ea, Guam\'a, Bel\'em, PA 66075-110, Brazil}


\author{A. A. Ara\'{u}jo Filho}
\email{dilto@fisica.ufc.br}
\affiliation{Departamento de Física, Universidade Federal da Paraíba, Caixa Postal 5008, 58051--970, João Pessoa, Paraíba,  Brazil.}
\affiliation{Departamento de Física, Universidade Federal de Campina Grande Caixa Postal 10071, 58429-900 Campina Grande, Paraíba, Brazil.}
\affiliation{Center for Theoretical Physics, Khazar University, 41 Mehseti Street, Baku, AZ-1096, Azerbaijan.}


\begin{abstract}

    In this work, we present a theoretical analysis of null geodesics, critical photon orbits, and shadow formation associated with a wormhole generated by a geometric defect. The propagation of light in this spacetime is examined through the deflection angle in both weak- and strong-field regimes. Analytical expansions are derived in each regime and employed to characterize gravitational lensing observables. By varying the global monopole charge, we evaluate its impact on these observables and determine parameter ranges that may be accessible to current or future observational probes. {Finally, we calculate the quasinormal modes and the time-domain solution for scalar perturbations.}

\end{abstract}
\maketitle

\tableofcontents


\section{Introduction}\label{sec1}

Singular black hole (BH) solutions emerge naturally within the framework of general relativity (GR) as a consequence of gravitational collapse, representing the final stage in the evolution of supermassive stars. The earliest and most fundamental theoretical description of such objects was introduced by Karl Schwarzschild, corresponding to a vacuum solution of Einstein’s field equations \cite{INTRO1,INTRO2,INTRO3,INTRO4}. Over time, several additional solutions were developed, including the electrically charged Reissner–Nordström spacetime associated with electrovacuum conditions, as well as models that incorporate a cosmological constant and rotating geometries, such as the Kerr and Kerr–Newman spacetimes \cite{INTRO1,INTRO2}. In order to eliminate the singularity present in the interior of a BH, the first regular BH model was proposed by J. M. Bardeen around 1968 \cite{INTRO5}. This spacetime does not satisfy the vacuum Einstein field equations \cite{INTRO5,INTRO6}, and for this reason its associated matter content was only identified around the year 2000 by Ayón-Beato and García \cite{INTRO7}. Such matter content is described by a class of nonlinear electrodynamics, in which the regularization parameter is interpreted as the charge of a magnetic monopole. The electrically charged version of the Bardeen spacetime was subsequently developed by Rodrigues and de S. Silva \cite{INTRO6}. Consequently, regular BH solutions have been investigated across a broad range of physical contexts \cite{INTRO8,INTRO9,INTRO10,INTRO11,INTRO12,INTRO13,INTRO14,INTRO15,INTRO16,INTRO17,INTRO18,INTRO19,INTRO20,INTRO21}.

Recently, further expanding the landscape of regular solutions, Simpson and Visser introduced a new class of solutions known as black bounces (BB) \cite{INTRO22}. In a simplified description, the main distinction of this model relative to a regular BH lies in the presence of an area function that can either conceal the singularity or, alternatively, describe a finite-radius throat associated with a wormhole (WH). In this model, different geometric configurations emerge according to the relationship between the mass parameter $m$ and the throat parameter $a$: for $a>2m$, a bidirectional WH is formed; for $a=2m$, a unidirectional WH is obtained with the throat located at the origin $r=0$; while for $a<2m$, a WH configuration with two symmetrical horizons arises. Analogously to Bardeen's model, BB spacetime, in its original formulation, does not constitute a solution to Einstein's equations. Only later did Bronnikov and Walia determine the material content corresponding to this spacetime, which is composed of a class of nonlinear electrodynamics coupled to a phantom scalar field \cite{INTRO23}. Since then, BB-type models have been extensively investigated in physical scenarios of interest \cite{INTRO24,INTRO25,INTRO26,INTRO27,INTRO28,INTRO29,INTRO30,INTRO31,INTRO32,INTRO33,INTRO34,INTRO35,INTRO36,INTRO37,INTRO38,INTRO39,INTRO40,INTRO41,INTRO42,INTRO43,INTRO44,INTRO45,INTRO45P}. All of these WH models presented above generally require, for their theoretical existence to be possible, some source of exotic matter that violates the energy conditions. In contrast, the model investigated in this work and proposed by the authors in Refs. \cite{B1,B2,B3} is a WH model created from a defect in spacetime \cite{B1}. We can also emphasize that these BB solutions have been explored in scenarios involving light deflection and massive particles, as well as in the study of gravitational lensing \cite{B5,B5A,B5B,B5C,B5D,B5D1,B5E,B5F,B5G,B5H}.

The phenomenon known as gravitational lensing consists of the deviation of light from a distant source as it passes through the gravitational field of a massive object, such as stars, BHs, and galaxies \cite{B5I,B5J}. In the specific case of BHs, it is possible to identify two distinct regimes. When light rays pass at relatively large distances from the compact object, the angle of deflection is small, characterizing the so-called weak-field regime, in which the deviation of light is moderate, and the images of the source do not show significant distortions. In contrast, as the light propagates closer and closer to the BH, approaching the photon sphere (region in which photons can describe stable circular orbits), the gravitational deflection becomes extremely intense, configuring the strong-field regime. In this context, photons can complete multiple orbits around the BH before reaching the observer, and the angle of deflection can even diverge, resulting in pronounced distortions. The images formed in this regime are called relativistic images. Pioneering work on the subject was proposed in the 1960s \cite{B5K,B5L}, which found that relativistic images were very weak and, therefore, there were no efficient mathematical techniques to calculate the deflection in the strong-field regime. It was only around the year 2000 that Virbhadra and Ellis \cite{B5M} developed a methodology for constructing minimally consistent gravitational lensing equations in the strong-field regime. Thus, Bozza \cite{B8} then developed a technique for obtaining the equations in the strong-field regime, which was later improved by Tsukamoto \cite{B9}. {Beyond the scenarios mentioned above, the study of gravitational lensing, deflection angle for both light and massive particles, shadows, and other possible observables has been explored for geometries with rotation and modified gravity \cite{EXTRA1,EXTRA2,EXTRA3,EXTRA4,EXTRA5,EXTRA6,EXTRA7,EXTRA8}.}

Among a range of models known as topological defects, there are those that arise via a mass generation process, known as the Higgs mechanism or spontaneous symmetry breaking. Some of these models, which possess very interesting characteristics, stand out, as those investigated in Minkowski (1+1)-dimensional spacetime and known as kinks. In two-dimensional Minkowskian spacetime (2+1), there are planar models such as vortices and skyrmions \cite{B9A,B9B,B9C,B9D,B9E}. These topological defects, when analyzed in the context of gravity, are generally called domain walls. A classic example of a domain wall is provided by the so-called cosmic strings that emerge in gravitational scenarios through the abelian Higgs model; they exhibit static and axial symmetry and maintain a direct connection with vortex solutions found in flat spacetime. As we know, the abelian Higgs model is associated with the spontaneous symmetry breaking mechanism, which is responsible for the mass generation of particles. In short, models that exhibit vortex-type solutions are used to characterize the superconducting properties of materials, allowing a correspondence between microscopic and macroscopic quantities and their eventual experimental verification \cite{B9A,B9C,B9E}. In recent years, there has been a growing number of research projects associated with understanding possible scenarios for building systems that can accommodate exact solutions and, naturally, analyze quantum dynamics in the global monopole spacetime. The investigation of the hydrogen atom, in both its relativistic \cite{B9F} and non-relativistic versions \cite{B9G}, stands out as a prime example. Furthermore, it has been analyzed to obtain bound states in Dirac and Klein-Gordon oscillators \cite{B9H}, as well as to construct WH-type solutions in Born-Infeld theory \cite{B5,B9J,B9K}.

After the coalescence of compact objects, the remnant spacetime undergoes a relaxation process that dominates the late portion of the emitted signal. When the highly dynamical and nonlinear regime subsides, the newly formed BH approaches equilibrium through a phase governed by damped oscillatory behavior. This regime is described by a discrete set of complex frequencies, the quasinormal modes, which arise as solutions of the linearized perturbation equations under appropriate boundary conditions~\cite{Konoplya:2007zx,Konoplya:2013rxa,karmakar2024quasinormal,Konoplya:2019hlu,Kokkotas:2010zd,Konoplya:2011qq}. These frequencies are not arbitrary: they are uniquely determined by the parameters characterizing the BH, including its mass and any additional charges or deformations present in the geometry and WHs \cite{Bronnikov:2021liv,batic2024unified,batic2025instability,batic2025spectral}. The temporal oscillations are controlled by the real part of each frequency, whereas the imaginary part sets the characteristic damping timescale. Since no information about the initial perturbation survives in this spectrum, any modification of the background spacetime manifests directly in the structure of the ringdown signal. For this reason, quasinormal modes have been explored in parallel with other geometric probes, such as BH shadows~\cite{Jusufi:2020dhz} and transmission properties associated with greybody factors~\cite{Konoplya:2024vuj,Konoplya:2024lir}. Claims that present gravitational--wave observations may already hint at individual quasinormal contributions remain under scrutiny, as their significance depends sensitively on statistical and systematic uncertainties~\cite{Franchini:2023eda}. Ongoing and future improvements in the sensitivity of the LIGO, Virgo, and KAGRA detectors are expected to sharpen these analyses as the catalog of detected mergers continues to grow.

{In contrast to previous investigations of the Schwarzschild–Klinkhamer WH, which focus on pure vacuum defects, this work addresses the gap in understanding how topological charges influence such geometries. Our results show that, although the throat radius a dominates the late-time stability behavior (QNMs), the global monopole charge  is the primary determining factor for both weak- and strong-field gravitational lensing observables, providing a unique physical decoupling between the BH size and its gravitational signatures.}

In Section \ref{sec2} of the present work, we present the general relations for a spherically symmetric metric and conserved quantities. In Section \ref{sec3}, we present the model; in Sections \ref{sec8} and \ref{sec8A} we analyze optical effects for the weak- and strong-field regimes. In Section \ref{sec4} we apply the gravitational lensing technique to the model and construct the observables. In Sections \ref{MQ} and \ref{TDOMAIN} we analyze quasinormal modes and time domain solutions. Finally, in Section \ref{sec6}, we present the final considerations and conclusions.


\section{General relations}\label{sec2}

As a starting point, we consider a static line element endowed with spherical symmetry, which characterizes the class of metrics depending solely on the radial coordinate and is widely employed in the description of stationary gravitational configurations. This line element can be expressed as:
\begin{equation}\label{1}
\mathrm{d}s^2= -f(r)\mathrm{d}t^2 + \frac{\mathrm{d}r^2}{g(r)f(r)} +\Sigma^2(r)\left(\mathrm{d}\theta^2+\sin^2\theta \,{\mathrm{d}\phi^2}\right),
\end{equation} where $f\left(r\right)$, $g\left(r\right)$, and $\Sigma\left(r\right)$ are functions that depend only on the radial coordinate. Thus, we define a smooth curve in this generic spacetime, Eq. \eqref{1}, that has length $S$ which is given by
\begin{equation}\label{2}
S= \int \sqrt{\left(g_{\mu\nu}\frac{\mathrm{d}x^\mu}{\mathrm{d}\lambda}\frac{\mathrm{d}x^\nu}{\mathrm{d}\lambda}\right)}\mathrm{d}\lambda,
\end{equation} where $\lambda$ is an affine parameter that can represent the observer's proper time. Taking $S$ as the affine parameter itself, we can show that the curve that minimizes Eq. (\ref{2}) is the same one that minimizes
\begin{equation}\label{3}
\int \left(g_{\mu\nu}\frac{\mathrm{d}x^\mu}{\mathrm{d}\lambda}\frac{\mathrm{d}x^\nu}{\mathrm{d}\lambda}\right)\mathrm{d}\lambda=\int\mathcal{L}\,{\mathrm{d}\lambda}.
\end{equation}

Therefore, considering the analysis in the equatorial plane, $\theta=\frac{\pi}{2}$, the Lagrangian becomes:
\begin{equation}\label{4}
\mathcal{L}= -f(r)\left(\frac{\mathrm{d}t}{\mathrm{d}\lambda}\right)^2 + \frac{1}{g(r)f(r)}\left(\frac{\mathrm{d}r}{\mathrm{d}\lambda}\right)^2 + \Sigma^2(r)\left(\frac{\mathrm{d}\phi}{\mathrm{d}\lambda}\right)^2.
\end{equation}

Applying the Euler-Lagrange equations to the above expression, we define the quantities conserved at time $t$ and at $\phi$. Therefore, we have to
\begin{equation}\label{5}
L= \Sigma^2(r)\frac{\mathrm{d}\phi}{\mathrm{d}\lambda},  \qquad \qquad E = f(r)\frac{\mathrm{d}t}{\mathrm{d}\lambda}.
\end{equation}

Thus, substituting the conserved quantities, Eq. \eqref{5}, in Eq. \eqref{4} and considering only null geodesics, $\mathcal{L}=0$, leads (\ref{4}) to
\begin{equation}\label{6}
\left(\frac{\mathrm{d}r}{\mathrm{d}\lambda}\right)^2= g(r)\left[E^2- \frac{L^2{f(r)}}{\Sigma^2(r)}\right].
\end{equation}

The expression above may be interpreted by analogy with the dynamics of a classical particle of unit mass, with energy $\mathcal{E}$, subject to an effective potential $V_{\text{eff}}$, given by
\begin{equation}\label{6A}
\mathcal{E}=E^2,\qquad\text{and}\qquad V_{\text{eff}}=\frac{L^2f(r)}{\Sigma^2} \ .
\end{equation}

Radial motion is permitted only when $\mathcal{E}>V_{\text{eff}}$. In this context, we examine the situation in which the photon begins its trajectory in an asymptotically flat region and approaches a distance 
$r_0$ from the center of BH, reaching the so-called turning point, located outside the event horizon. As expected, once the photon reaches this limiting region, the gravitational field causes it to return toward another asymptotically flat region. At the turning point, one has, $V_{\text{eff}}(r_0)=E^2$, which, from Eq. (\ref{6A}), leads to
\begin{equation}\label{6B}
	\frac{1}{\beta^2}=\frac{f(r_0)}{\Sigma^2 (r_0)} \ ,
\end{equation}
where $\beta=\frac{L}{E}$ is the  impact parameter of the light ray. At the photon sphere radius, $r_m$, we have
\begin{equation}\label{6C}
	\frac{\mathrm{d}V_{\text{eff}}}{\mathrm{d}r}\Bigg|_{r_m}=0 \ .
\end{equation}

The critical impact parameter, $\beta_c$, is defined by the condition 
$\beta_c=\beta(r_m)$. Light rays with $\beta<\beta_c$ are completely absorbed, whereas those for which $\beta=\beta_c$ remain trapped on the photon sphere; in contrast, rays with $\beta>\beta_c$ are deflected and ultimately scattered. In the scattering regime, the deflection angle diverges in the limit $r_0 \to{r_m}$, which characterizes the so-called strong-field regime. In the following sections, we derive the approximation for the deflection angle in these two regimes, considering the Schwarzschild–Klinkhamer WH immersed in the field of a global monopole.


\section{The Model}\label{sec3}

We will apply the methodology used for calculating the deflection of light when subjected to the effects of a gravitational field in a spherically symmetric and static configuration to a Schwarzschild-Klinkhamer WH topologically charged due to the presence of the global monopole and originally derived in Refs. \cite{B1,B2,B3,B3P} and subsequently the insertion of some types of topological defects in Refs. \cite{B7,B7P,ahmed2024five} was investigated. The central idea of this spacetime is to construct a model that describes a traversable WH without requiring exotic matter, which in GR is typically associated with the violation of the energy conditions. In contrast, in the model under consideration, such a solution becomes feasible by introducing a geometric defect \cite{B1}. The scenario containing multiple vacuum solutions was investigated in Ref. \cite{B2} and subsequently explored for the context of the generalized Schwarzschild and then investigated energy conditions, geodesics and some optical properties in Ref. \cite{B3}.

Thus, considering the spherically symmetric spacetime described by the line element in the form Eq. (\ref{1}), we have that the metric functions are defined as
\begin{equation}\label{7}
\Sigma^2(r)= r^2+a^2, \qquad \qquad f(r)= 1-\frac{2M}{\Sigma(r)}, \qquad \mbox{and} \qquad g(r)= \frac{\bar{\alpha}^2\Sigma^2(r)}{r^2},
\end{equation} where the parameters $M$ and $a$ are, respectively, the mass and radius of the WH throat and $\bar{\alpha}$ is associated with the charge of the global monopole. The validity regime of the metric tensor coordinates is defined by: the coordinates $t$ and $r$ associated with the line element Eq. (\ref{1}) and defined in Eq. (\ref{7}) $\in$ to the interval $(-\infty,\infty)$, $\theta$ $\in$ $[0,\pi)$ and $\phi$ $\in$ $[0,2\pi)$. {To represent a traversable WH, the relation $a>2M$ must be fixed, similarly to what occurs in the Simpson-Visser model \cite{INTRO22}. The validity regime of the parameter associated with the global monopole charge is defined in the range $0<\bar{\alpha}\leq1$.}

The event horizon of this model is located at $r_h=\pm\sqrt{(2M)^2-a^2}$ and, using Eq. (\ref{6C}), we have that the photon sphere is located at $r_m=\pm\sqrt{(3M)^2-a^2}$. Adopting the symmetry of the system, in the following sections we will only use the positive sign of the quantities defined here.


\section{\label{sec8}Light Propagation}

In this section, we examine the propagation of light in the WH geometry. The analysis begins with a study of null geodesics, obtained by solving a system of four coupled differential equations that govern photon trajectories. These solutions are then used to construct the corresponding geodesic plots. Subsequently, the radii of the photon sphere and the associated shadow are determined.


\subsection{Geodesics}

This section examines the motion of test particles by reconstructing their trajectories directly from the geometric structure of the spacetime. The analysis is organized around the spacetime connection, as the affine structure governs how worldlines deviate and bend. The starting point, therefore, is the explicit construction of the Christoffel symbols associated with the metric. Once these coefficients are established, the equations governing particle motion follow in a systematic way. For this reason, we first introduce the relevant expressions by writing:
\ie
\frac{\mathrm{d}^{2}x^{\mu}}{\mathrm{d}\lambda^{2}} + \Gamma\indices{^\mu_\alpha_\beta}\frac{\mathrm{d}x^{\alpha}}{\mathrm{d}\lambda}\frac{\mathrm{d}x^{\beta}}{\mathrm{d}\lambda} = 0. \label{geogeo}
\fe

Within the adopted framework, the evolution of a particle trajectory is parametrized by the affine parameter $\lambda$, which labels successive points along the worldline. When this parametrization is implemented in the equations of motion, the dynamics unfolds as a coupled system of four differential relations. Each equation governs the evolution of a specific spacetime coordinate, and none of them can be treated independently of the others. The complete set of equations is therefore expressed in the form:
\ie
\frac{\mathrm{d^2{t}}}{\mathrm{d\lambda^2}}=\Ddot{t} = -\frac{2 M r\dot{r} \dot{t}}{\left(a^2+r^2\right) \left(\sqrt{a^2+r^2}-2 M\right)},
\fe
\ie
\begin{split}
 \frac{\mathrm{d^2}r}{\mathrm{d}\lambda^2} =\Ddot{r}=&  \frac{\Bar{\alpha}^2 \left(-2 M \sqrt{a^2+r^2}+a^2+r^2\right) \left(\left(a^2+r^2\right)^{3/2} \left(\left(\dot{\theta}\right)^2+\sin ^2(\theta ) \left(\dot{\varphi}\right)^2\right)-M \left(\dot{t}\right)^2\right)}{r^{2}\left(a^2+r^2\right)^{3/2}}\\
& -\frac{\left(\dot{r}\right)^2 \left(\frac{M r^2}{2 M-\sqrt{a^2+r^2}}+a^2\right)}{r^{2}(a^2+r^2)},
\end{split}
\fe
\ie
\begin{split}
& \frac{\mathrm{d^2} \theta}{\mathrm{d} \lambda^2}=\Ddot{\theta} =  \sin (\theta ) \cos (\theta ) \left(\dot{\varphi}\right)^2-\frac{2 r \dot{\theta}\dot{r}}{a^2+r^2},
\end{split}
\fe
and 
\ie
\begin{split}
& \frac{\mathrm{d^2} \varphi}{\mathrm{d} \lambda^2}=\Ddot{\varphi}  = 2 \dot{\varphi} \left(-\frac{r\dot{r}}{a^2+r^2}-\dot{\theta} \cot (\theta )\right),
\end{split}
\fe
where the dot symbol $(\,^{\dot{}}\,)$ denotes differentiation with respect to the affine parameter $\lambda$.

\begin{figure}
    \centering
    \includegraphics[scale=0.63]{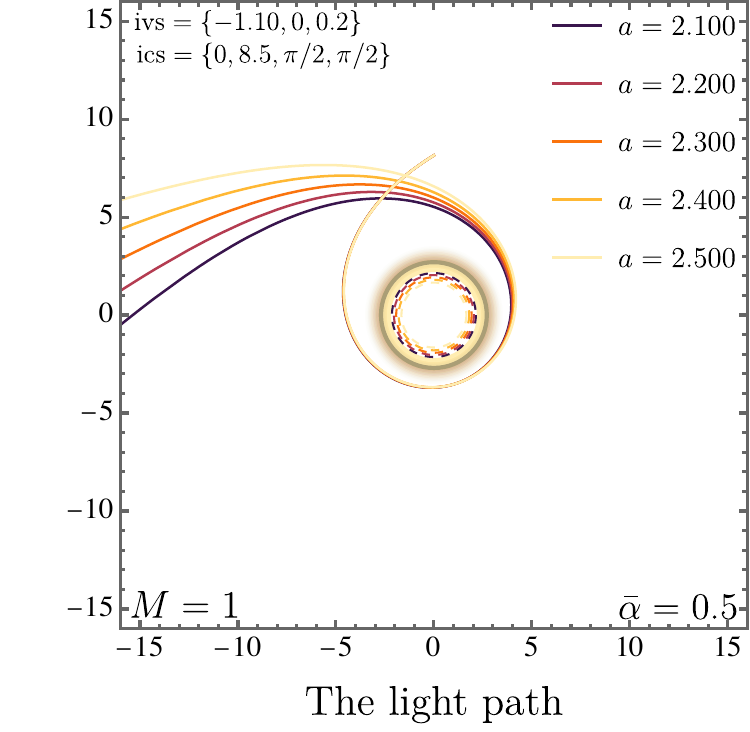}
    \includegraphics[scale=0.63]{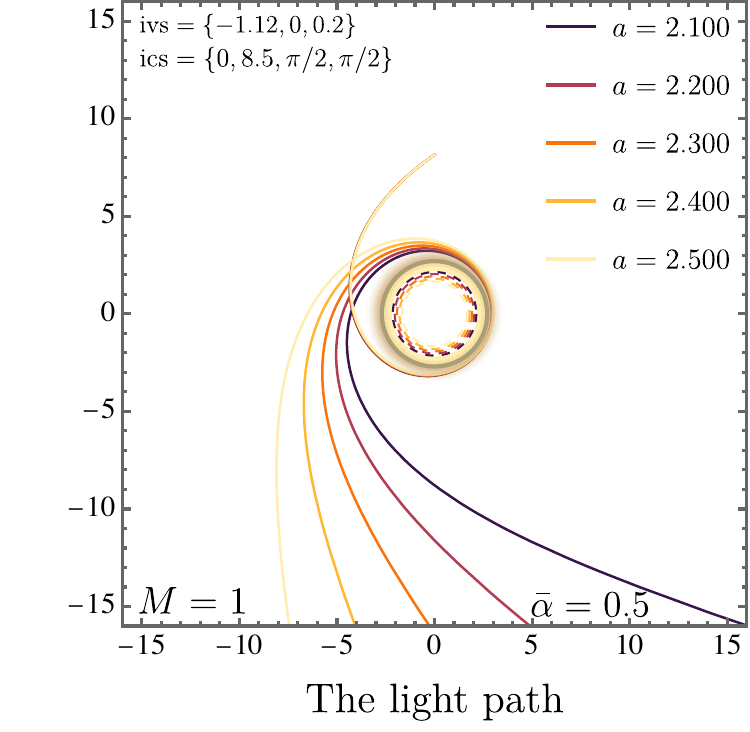}
    \includegraphics[scale=0.63]{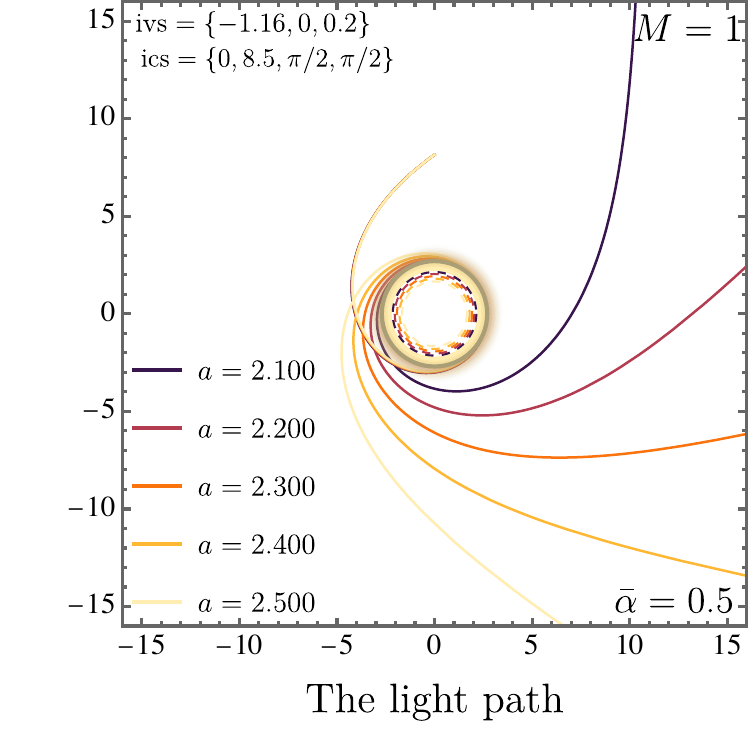}
    \includegraphics[scale=0.63]{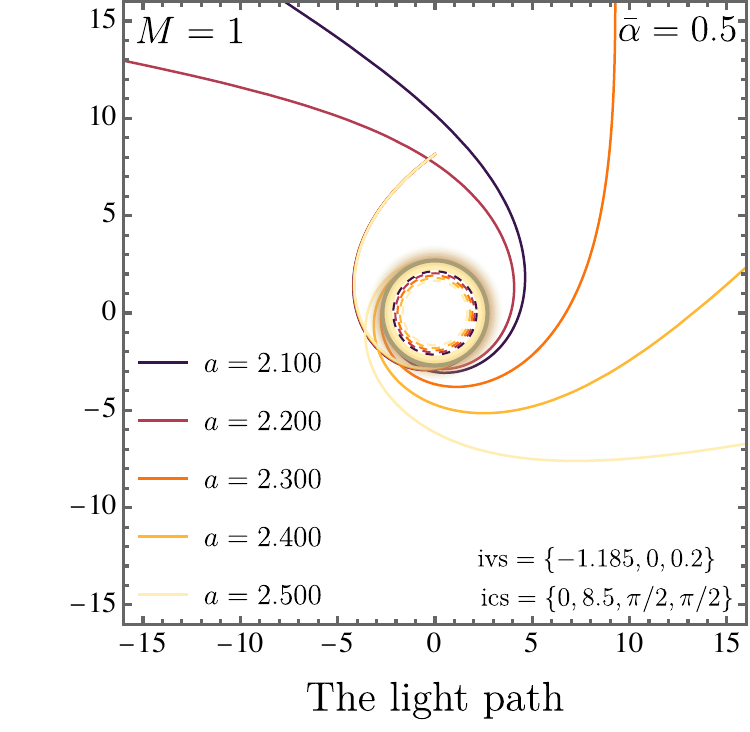}
    \caption{ {Null geodesics for fixed $\Bar{\alpha}=0.5$ and $M=1$, with different values of $a$. The four panels correspond to distinct initial data. The labels $ics$ and $ivs$ denote the initial coordinates and initial derivatives, respectively, while $\dot t(0)$ is determined by the null condition. The dashed curves represent the associated photon spheres. The axes are given in units of $M$.} }
    \label{geooooo2}
\end{figure}

{Figure~\ref{geooooo2} displays null geodesics for a fixed value of $\Bar{\alpha}=0.5$, with the mass parameter set to $M=1$, while the parameter $a$ is varied in each panel. The four panels correspond to different choices of initial data for the light rays, which explains the distinct incidence directions shown in the plots. In our notation, $ics={t(0),r(0),\theta(0),\phi(0)}$ denotes the initial position, whereas $ivs={\dot r(0),\dot\theta(0),\dot\phi(0)}$ represents the freely prescribed initial derivatives; the quantity $\dot t(0)$ is then determined from the null condition. For fixed $\Bar{\alpha}$, increasing $a$ makes the corresponding trajectories progressively more open, indicating a weaker deflection of the light path. The dashed curves represent the corresponding photon spheres, whose radii change with $a$ while $\Bar{\alpha}$ remains fixed. The horizontal and vertical axes are given in units of $M$.}

\subsection{Critical orbits and shadows radii}\label{sec:null}

The behavior of light rays is analyzed by constructing an effective description based on the Lagrangian approach, as discussed before. {Within this framework, the equations governing the motion of photons are derived directly from a variational principle applied to the geometry of spacetime, resulting in the following expression $\mathcal{L} = \frac{1}{2}{g_{\mu \nu }}{{\dot x}^\mu }{{\dot x}^\nu }$, which, when we use metric functions, becomes:}

\begin{equation}
\label{lagrangian}
\mathcal{L} = \frac{1}{2}\Big[ - A(r){{\dot t}^2} + B(r){{\dot r}^2} + C(r){{\dot \theta }^2} + D(r){{\mathop{\rm \sin}\nolimits} ^2}\, \theta {{\dot \varphi }^2}\Big].
\end{equation}
Implementing the Euler–Lagrange equations and restricting the motion to the equatorial section of the geometry ($\theta=\pi/2$), two constants of motion emerge as a direct consequence of the underlying symmetries. These correspond to the conserved energy $E=A(r)\dot t$ and angular momentum $L= D(r)\dot \varphi$ associated with photon propagation. 
{By substituting the conserved quantities into the equation of motion and performing the necessary algebraic manipulations, the equation governing the system can be rewritten in the form:}

\begin{equation}\label{rdot}
\frac{{{{\dot r}^2}}}{{{{\dot \varphi }^2}}} = {\left(\frac{{\mathrm{d}r}}{{\mathrm{d}\varphi }}\right)^2} = \frac{{D(r)}}{{B(r)}}\left(\frac{{D(r)}}{{A(r)}}\frac{{{E^2}}}{{{L^2}}} - 1\right)=\mathrm{V}_{\text{eff}}(r),
\end{equation} {where $\mathrm{V}_{\text{eff}}(r)$ is the effective potential of the system. For further verification on how to derive equations in general, see Ref. \cite{B7}.}


Once the preceding elements are in place, the characterization of the photon spheres can be addressed. As discussed previously, their determination follows from enforcing the appropriate extremal condition on the radial motion, which is imposed as
$\mathrm{V}_{\text{eff}}(r)=E^2$ and $ \frac{\mathrm{d} \,{\mathrm{V}_{\text{eff}}(r)}}{\mathrm{d}r} = 0$.

Recalling that we had defined the critical impact parameter as $\beta_c=L/E$, the first of these conditions directly yields
\ie
\beta_c=\sqrt{\frac{D(r)}{A(r)}}\Bigg|_{r=r_{m}}.
\fe

The analysis proceeds by resolving the equation obtained above. Its resolution reveals a unique solution that is both real and positive, which is naturally identified with the radius of the photon sphere and denoted by $r_{m}$. The explicit expression, as we said before, is given by:
\ie
\label{crittiaal}
r_{m} = \sqrt{9 M^2-a^2}.
\fe
Notice that in the limit $a \to 0$, the contribution vanished and the photon sphere radius of the Schwarzschild geometry was exactly recovered. Notably, the topological charge $\Bar{\alpha}$ does not enter the expression and therefore produces no correction at this order.

\begin{table}[!ht]
   \centering
    \caption{{Numerical values of the photon sphere radius $r_{m}$ for different values of the parameter $a$, with the BH mass fixed to $M=1$. }} \begin{tabular}{|c|c|c|c|c|c|c|}
    \hline
         $r_{m}$ & $a = 0.0$ & $a = 2.1$ & $a = 2.2$ & $a = 2.3$& $a = 2.4$ & $a = 2.5$
         \\ \hline\hline
        ------ & 3.00000 & 2.14243 & 2.03961 & 1.92614 & 1.80000 & 1.65831 \\ \hline
    \end{tabular}
    \label{Tab:rphoton}
\end{table}

Table~\ref{Tab:rphoton} compiles the computed values of the photon sphere radius $r_{m}$ for a BH of unit mass, $M=1$, considering different values of the parameter $a$. The numerical results reveal a clear trend: as $a$ is increased, the location of the photon sphere shifts inward, leading to a smaller value of $r_{m}$. Furthermore, following the standard construction adopted in Refs.~\cite{perlick2015influence,konoplya2019shadow,AraujoFilho:2025zaj,AraujoFilho:2025hnf,AraujoFilho:2024lsi}, the angular radius of the shadow cast by a spherically symmetric BH can be expressed in the form
\ie
\label{shadow}
\begin{split}
 & R_{sh}  =   \sqrt {\frac{{{D(r)(r_{m})}}}{{{A(r)(r_{m})}}}} = 3 \sqrt{3} M.
\end{split}
\fe
In this expression, the complete form of the photon sphere radius has been adopted and the analysis has been confined to the equatorial plane ($\theta=\pi/2$). Under these conditions, no explicit contribution from either the parameter $a$ or the topological charge $\Bar{\alpha}$ appears in the shadow radius.


\section{\label{sec8A}Lensing phenomena}

    
\subsection{Expansion for light  deflection in the weak-field limit}\label{sec31}

Considering the light deflection in this specific scenario, where the metric functions are given by Eq. (\ref{1}), one obtains the following expression for the impact parameter of the light ray:
\begin{equation}\label{8}
    \frac{1}{\beta^2(r_0)}= \frac{\sqrt{r^2_0+a^2}-2M}{(r^2_0+a^2)^{3/2}}.
\end{equation}
Using Eq. (\ref{7}), the radial equation, Eq. (\ref{6}), becomes
\begin{equation}\label{9}
	\bigg(\frac{\mathrm{d}r}{\mathrm{d}\lambda}\bigg)^2= \frac{\bar{\alpha}^2\Sigma^2}{r^2}\left[E^2-\frac{L^2f(r)}{\Sigma^2} \right].
\end{equation}
Substituting Eq. (\ref{5}) in Eq. (\ref{9}), we find
\begin{equation}\label{10}
	\Big(\frac{\mathrm{d}r}{\mathrm{d}\phi}\Big)^2= \frac{\bar{\alpha}^2\Sigma^6}{L^2r^2}\left[E^2-\frac{L^2f(r)}{\Sigma^2} \right] .
\end{equation}

Given that the distances before and after the inflection point are symmetrical and therefore the contributions to the angular deviation are the same, we then have the expression for the angular deviation:
\begin{equation}\label{11}
\Delta\phi= \pm{\frac{2L}{\bar{\alpha}}}\int^{\infty}_{r_0}\frac{r}{\Sigma^3}\left[E^2-\frac{L^2{f(r)}}{\Sigma^2}\right]^{-1/2}{\mathrm{d}r},
\end{equation} in developing this work, for the sake of symmetry, we only adopted the positive sign from the expression above.

We now introduce a coordinate transformation in Eq. (\ref{11}), where $u=\frac{1}{\sqrt{r^2+a^2}}$. Note that, under this transformation, the integration limits become $u\to{0}$ as $r\to\infty$, and $u\to{u_0}=\frac{1}{\sqrt{r^2_0+a^2}}$ as $r\to{r_0}$. Consequently, the equation for the angular deviation in the new coordinates takes the form
\begin{equation}\label{12}
    \Delta\phi= \frac{2}{\bar{\alpha}}\int^{u_0}_0{\mathrm{d}u}\left[\frac{1}{\beta^2(r_0)}-u^2+2Mu^3\right]^{-1/2}.
\end{equation}

To carry out the expansion of the angular deviation in the weak-field regime, we must return to the expression for the impact parameter given in Eq. (\ref{8}) and then apply the same coordinate transformation introduced above, $\frac{1}{\beta^{2}} = u_{0}^{2}\left(1 - {2M}u_{0}\right)$. Using this new form of the impact parameter and assuming that the photon travels far from the WH, we may adopt the approximation of a small mass $M \ll 1$. Consequently, by expanding the orbital equation, Eq. (\ref{12}), up to the second order in $M$, we find that the deflection of light is given by $\delta\phi = \Delta\phi - \pi$
\begin{equation}\label{13}
\delta\phi \simeq \frac{1}{\bar{\alpha}}\Bigg[\pi(1-\bar{\alpha})+\frac{4M}{\beta} +\frac{15\pi M^2}{4\beta^2}\Bigg].
\end{equation}

In the expression for the angular deviation for the weak-field regime given above, we can observe that it does not depend on the parameter that describes the area function $a$. On the other hand, the deviation depends only on the mass parameter $M$ and the parameter associated with the global monopole charge $\bar{\alpha}$.

We can also observe in the expression referring to the angular deviation Eq. (\ref{13}) that in the limit where the monopole charge ceases to act $\bar{\alpha}\to{1}$ we recover the result for the scenario of light deflection in the Schwarzschild BH background with mass correction up to second order \cite{B5}. On the other hand, when considering the scenario where the mass of the solution is zero $M\to{0}$, we recover the angular deviation in the spacetime of the global monopole, in agreement with the results obtained in Refs. \cite{B9K,B7}.

\begin{figure}[htb!]
\centering
	{\includegraphics[width=0.95\textwidth]{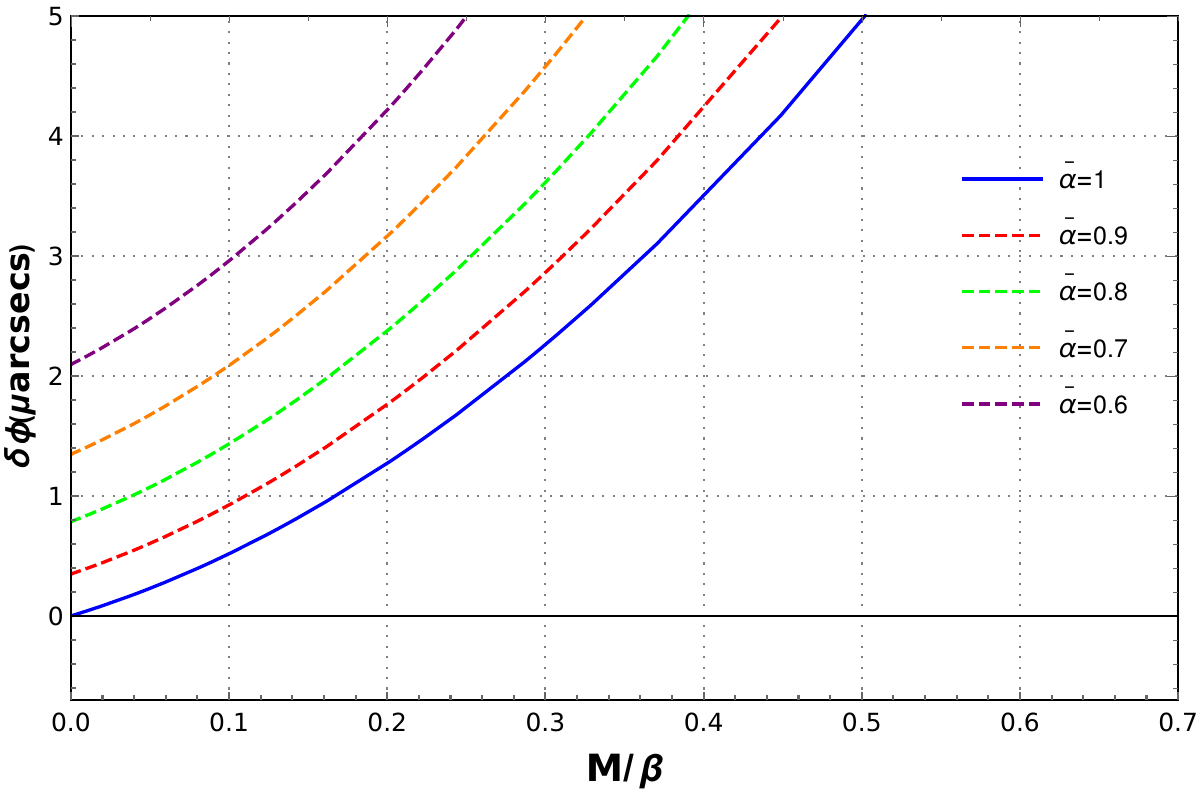}}
    \caption{Light deflection in the weak-field regime.}
\label{FRACO}
\end{figure}

In Fig. (\ref{FRACO}) we have the representation of the light deviation in the weak-field regime as a function of the ratio $M/\beta$, taking into account the variation of the parameter associated with the charge of the global monopole. The solid curve represents that the monopole's charge is {switched off}, so we exclusively have the Schwarzschild BH, and the dashed curves represent how the charge on the global monopole influences the deflection angle of the BH.


\subsection{The Gauss--Bonnet approach}

This section is devoted to the study of light bending in the weak-deflection limit. The calculation of the deflection angle is performed within the geometric framework provided by the Gauss--Bonnet approach \cite{Gibbons:2008rj}, which relates global properties of the optical manifold to the bending of null trajectories.

As a preliminary step, the character of the photon spheres introduced in Eq.~(\ref{crittiaal}) is examined from a stability perspective. This assessment is carried out through the evaluation of the Gaussian curvature associated with the corresponding optical geometry. The curvature is central to this analysis: its sign directly classifies the nature of the critical trajectories, with positive values signaling stability and negative values indicating unstable photon orbits.

For pedagogical purposes, let us start from Eqs.~\eqref{1} and \eqref{7}.
Substituting
$g(r)=\bar{\alpha}^{2}\Sigma^{2}(r)/r^{2}$
and
$\Sigma^{2}(r)=r^{2}+a^{2}$,
we obtain
\begin{equation}
\mathrm{d}s^{2}=
-f(r)\,\mathrm{d}t^{2}
+\frac{r^{2}}{\bar{\alpha}^{2}\Sigma^{2}(r)\,f(r)}\,\mathrm{d}r^{2}
+\Sigma^{2}(r)\,\mathrm{d}\Omega^{2},
\qquad
f(r)=1-\frac{2M}{\Sigma(r)}.
\nonumber
\end{equation}

To rewrite the line element in a Schwarzschild--like form, we introduce the
areal radius
\begin{equation}
R \equiv \Sigma(r)=\sqrt{r^{2}+a^{2}},
\qquad\Rightarrow\qquad
\mathrm{d}R=\frac{r}{R}\,\mathrm{d}r,
\ \ \ \mathrm{d}r^{2}=\frac{R^{2}}{r^{2}}\,\mathrm{d}R^{2}.
\end{equation}
Then the radial sector becomes
\begin{equation}
\frac{r^{2}}{\bar{\alpha}^{2} R^{2} f(R)}\,\mathrm{d}r^{2}
=
\frac{r^{2}}{\bar{\alpha}^{2} R^{2} f(R)}
\left(\frac{R^{2}}{r^{2}}\,\mathrm{d}R^{2}\right)
=
\frac{1}{\bar{\alpha}^{2} f(R)}\,\mathrm{d}R^{2},
\end{equation}
where we used $\mathrm{d}R=(r/R)\,\mathrm{d}r$ and therefore
\begin{equation}
f(R)=1-\frac{2M}{R}.
\end{equation}
In other words, we have
\begin{equation}
\mathrm{d}s^{2}= 
-\left(1-\frac{2M}{R}\right)\mathrm{d}t^{2}
+
\frac{1}{\bar{\alpha}^{2}\left(1-\frac{2M}{R}\right)}\,\mathrm{d}R^{2}
+
R^{2}\,\mathrm{d}\Omega^{2}.
\end{equation}

Now perform the constant rescaling of the radial coordinate
\begin{equation}
\tilde{r}\equiv \frac{R}{\bar{\alpha}}
=\frac{\sqrt{r^{2}+a^{2}}}{\bar{\alpha}},
\qquad
\Rightarrow\qquad
R=\bar{\alpha}\tilde{r},
\ \ \ 
\mathrm{d}R=\bar{\alpha}\,\mathrm{d}\tilde{r},
\end{equation}
and redefine the mass parameter and the angular prefactor as
\begin{equation}
\label{reducedddd}
\tilde{M}\equiv \frac{M}{\bar{\alpha}},
\qquad
\tilde{K}^{2}\equiv \bar{\alpha}^{2}.
\end{equation}
The metric can then be written as
\begin{equation}\label{redefinedmetricWH}
\mathrm{d}s^{2}=
-\left(1-\frac{2\tilde{M}}{\tilde{r}}\right)\mathrm{d}t^{2}
+\frac{1}{1-\frac{2\tilde{M}}{\tilde{r}}}\,\mathrm{d}\tilde{r}^{\,2}
+\tilde{K}^{2}\tilde{r}^{2}\,\mathrm{d}\Omega^{2},
\end{equation}
which preserves the Schwarzschild functional form in the $(t,\tilde{r})$
sector, while the global--monopole--type modification is encoded in the
constant factor $\tilde{K}^{2}$ multiplying the angular term (note that
$\tilde{r}$ is not the areal radius, since $R=\bar{\alpha}\tilde{r}$).

In this case, the application of the Gauss--Bonnet technique requires the
inclusion of additional elements not present in the original formulation of
Ref.~\cite{Gibbons:2008rj}, following the procedures adopted in
Refs.~\cite{Jusufi:2017lsl,AraujoFilho:2025zaj}.

\subsubsection{Stability of the critical orbits }

Photon rings, also referred to as critical light orbits, emerge as a consequence of the geometric structure encoded in the optical manifold associated with a spacetime. Their dynamical behavior is not fixed by symmetry alone but is instead controlled by how nearby null trajectories respond to small deviations from circular motion. Exact circular paths are therefore idealized limits: once perturbed, photons either drift away from the orbit or remain confined in its vicinity, depending on the underlying geometric conditions. In configurations associated with instability, even infinitesimal disturbances drive photons toward the horizon or allow them to escape to infinity, whereas in stable settings the trajectories linger near the original radius, executing repeated revolutions around the BH \cite{qiao2022geometric,Heidari:2025iiv,qiao2022curvatures,araujo2025impact}.

This behavior admits a natural reformulation in geometric terms by examining the intrinsic curvature of the optical manifold. In this description, the Gaussian curvature $\mathcal{K}(r)$ governs the relative behavior of neighboring light rays along their propagation. Regions characterized by nonpositive curvature, $\mathcal{K}(r)\leq 0$, are excluded from supporting conjugate points according to the Cartan--Hadamard theorem, which implies that circular photon trajectories in such regions cannot be stable. By contrast, when $\mathcal{K}(r)>0$, conjugate points may appear, allowing for the existence of localized photon rings that resist small perturbations \cite{qiao2024existence}. From this viewpoint {and to facilitate the reproducibility and traceability of the results for a generic metric tensor, we employ the general definition presented in Eq.~(\ref{lagrangian}), which depends only on the metric components. Then, the} null trajectories satisfying $\mathrm{d}s^{2}=0$ can be recast in an equivalent geometric form, which we write as \cite{AraujoFilho:2024xhm}:
\begin{equation}
\mathrm{d}t^2 = \Tilde{\gamma}_{ij}\mathrm{d}x^i \mathrm{d}x^j = \frac{1}{A(r) \, B(r)}\mathrm{d}r^2  +\frac{\Bar{D}(r)}{A(r)}\mathrm{d}\varphi^2.
\end{equation}

Within this construction, the labels $i$ and $j$ refer exclusively to the three spatial directions, taking values from $1$ to $3$, while $\tilde{\gamma}_{ij}$ represents the components of the induced metric on the optical geometry. The function $\bar{D}(r)$ is defined by evaluating the original metric function on the equatorial section, that is, $\bar{D}(r)\equiv D(r,\theta=\pi/2)$. Once these elements are specified, the geometric content of the optical manifold is entirely encoded in its Gaussian curvature. An explicit expression for this curvature, derived in Ref.~\cite{qiao2024existence}, is given by
\begin{equation}
\mathcal{K}(r) = \frac{R}{2} =  -\frac{ A(r) \sqrt{B(r)}}{\sqrt{ \,  \Bar{D}(r)}}  \frac{\partial}{\partial r} \left[  \frac{A(r) \sqrt{B(r)}}{2 \sqrt{  \Bar{D}(r) }}   \frac{\mathrm{d}}{\mathrm{d} r} \left(   \frac{\Bar{D}(r)}{A(r)}    \right)    \right].
\end{equation}
In this context, $R$ denotes the Ricci scalar associated with the two--dimensional optical metric. When the parameter $a$ is assumed to be small, the Gaussian curvature admits a perturbative expansion in powers of $a$, leading to the following approximate form:
\begin{equation}
\begin{split}
\label{gbbhhh}
& \mathcal{K}(r) = \frac{3 \Tilde{M}^2}{\Tilde{r}^4}-\frac{2 \Tilde{M}}{\Tilde{r}^3} = \frac{3 \bar{\alpha} ^2 M^2}{\left(a^2+r^2\right)^2}-\frac{2 \bar{\alpha} ^2 M}{\left(a^2+r^2\right)^{3/2}}.
\end{split}
\end{equation}

Earlier investigations have established that the fate of circular photon trajectories under small disturbances is controlled by the Gaussian curvature $\mathcal{K}(r)$ of the optical manifold \cite{Heidari:2025iiv,araujo2025impact,qiao2024existence,qiao2022geometric,AraujoFilho:2025huk,qiao2022curvatures,AraujoFilho:2025vgb}. The sign of this curvature determines the local behavior of neighboring null geodesics. When $\mathcal{K}$ is positive, nearby light rays tend to focus, allowing closed photon paths to remain confined. In contrast, negative curvature leads to an increasing separation between neighboring trajectories, rendering circular orbits unstable and causing photons to either fall into the compact object or propagate outward to infinity.

\begin{figure}
    \centering
    \includegraphics[scale=0.68]{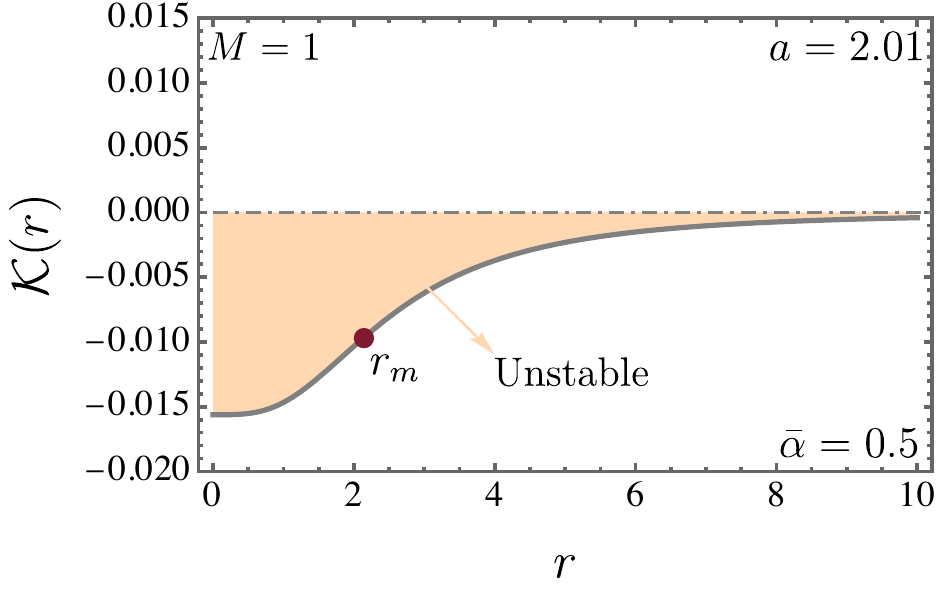}
    \caption{The Gaussian curvature $\mathcal{K}(r)$ evaluated for $M=1$, $\Bar{\alpha} =0.5$ and $a=2.01$. The solid wine marker identifies the radial location at which $\mathcal{K}$ vanishes, defining the boundary between regions of stability and instability for photon motion. The wine dotted marker denotes the photon sphere radius $r_{m}$, which lies entirely within the unstable domain. }
    \label{cuvgahjj}
\end{figure}

This behavior is illustrated in Fig.~\ref{cuvgahjj}, where $\mathcal{K}(r)$ is displayed as a function of the radial coordinate for the representative choice $M=1$, $\Bar{\alpha}$, and $a=2.01$. Only a single region was identified: an instability domain, highlighted in light orange. Within this sector, no stable orbital configurations existed. Because the photon sphere radius lay entirely inside the orange region, every circular photon orbit admitted by this spacetime belonged to the unstable regime.

\subsubsection{Weak deflection angle }

The bending of light in the weak-field regime is evaluated through a geometric procedure based on the Gauss--Bonnet theorem \cite{Gibbons:2008rj}. The calculation makes direct use of the curvature given in Eq.~(\ref{gbbhhh}). For practical implementation, the optical geometry is projected onto the equatorial section by fixing $\theta=\pi/2$, which effectively yields a two--dimensional manifold. Within this setting, the associated surface element assumes the form:
\begin{equation}
\mathrm{d}S = \sqrt{\Tilde{\gamma}} \, \mathrm{d} r \mathrm{d}\varphi = \sqrt{\frac{1}{A(r)} \frac{1}{B(r)} \frac{D(r)}{A(r)} } \, \mathrm{d} r \mathrm{d}\varphi.
\end{equation}

The evaluation of the integral leading to the deflection angle is carried out under assumptions that are compatible with the geometric approach adopted here. In particular, the impact parameter is taken in the asymptotic regime $\beta \gg 2M$, in accordance with the original application of the Gauss–Bonnet theorem to gravitational lensing \cite{Gibbons:2008rj}. The treatment of the parameter $a$ follows the same perturbative strategy used throughout this analysis, while the mass dependence is retained up to second order. This truncation scheme is standard in weak-field lensing calculations and has been employed extensively in related investigations \cite{araujo2024effects,AraujoFilho:2025huk,AraujoFilho:2025hnf,araujo2025geodesics}.

\begin{figure}
    \centering
    \includegraphics[scale=0.6]{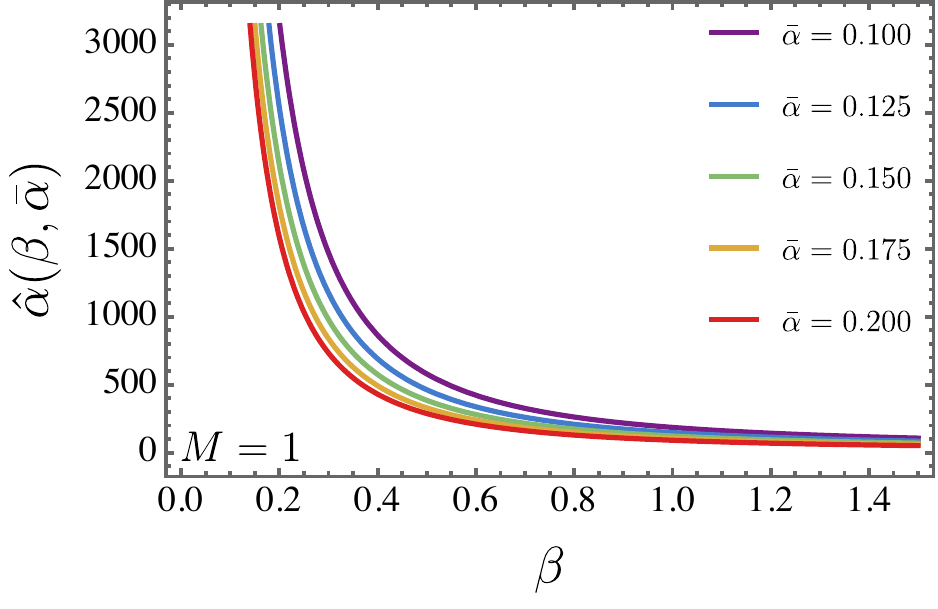}
    \caption{The deflection angle $\hat{\alpha}(\beta,\Bar{\alpha})$ is plotted as a function of the impact parameter $\beta$ for several values of  $\Bar{\alpha}$.}
    \label{cuuggsss}
\end{figure}

Starting from the expression introduced above, the resulting formula for the light--deflection angle can then be written as \cite{Jusufi:2017lsl}
\begin{equation}
\begin{split}
\label{vfffdddd}
& \hat{\alpha} (\beta,\Bar{\alpha}) =  \pi \left( \frac{1}{\Tilde{K}} -1 \right) - \frac{1}{\Tilde{K}}\int^{\pi}_{0} \int^{\infty}_{\big(\frac{\sin (\varphi )}{\Tilde{\beta}}+\frac{\Tilde{M} (1-\cos (\varphi ))^2}{ \Tilde{\beta}^2}\big)^{-1}} \mathcal{K} \mathrm{d}S \\
& \simeq  \, \pi  \left(\frac{1}{\Bar{\alpha} }-1\right)+ \frac{1}{\Bar{\alpha} } \left(\frac{4  \Tilde{M}}{\Tilde{\beta}}  + \frac{15 \pi   \Tilde{M}^2}{4 \Tilde{\beta}^2}\right) \\
& = \pi  \left(\frac{1}{\Bar{\alpha} }-1\right)+ \frac{1}{\Bar{\alpha} } \left(\frac{4 M}{\beta}  + \frac{15 \pi M^2}{4 \beta^2}\right),
\end{split}
\end{equation}
{where $\tilde{\beta}$ denotes the modified impact parameter induced by the redefinition of the original metric in Eq.~(\ref{1}) into the monopole-like form given in Eq.~(\ref{reducedddd}). It is related to the original impact parameter $b$ by $\Tilde{\beta} = \beta/\bar{\alpha}$.} A simple but important point should be emphasized: in the simultaneous limits $\Bar{\alpha}\to 1$ and $a\to 0$, the standard Schwarzschild configuration is recovered. {This result is the same obtained in Eq. (\ref{13})}. In addition, Fig.~\ref{cuuggsss} shows the behavior of the deflection angle $\hat{\alpha}(\beta,\Bar{\alpha})$ as a function of the impact parameter $\beta$ for different values of $\Bar{\alpha}$. All curves are obtained by fixing the mass parameter to $M=1$. The results larger values of the topological charge $\Bar{\alpha}$ produce an overall reduction of $\hat{\alpha}(\beta,\Bar{\alpha})$.

\subsection{Deflection of light in the strong-field limit}\label{sec32}

To calculate the gravitational deflection of light in the strong-field limit, we use as a basis the protocol originally established by Bozza \cite{B8} and later improved by Tsukamoto \cite{B9}. In this context, we begin by transforming the variables $z=1-\frac{r_0}{r}$ in the orbit equation, Eq. (\ref{11}), from which it can be expressed again as:
\begin{equation}\label{14}
\Delta\phi(r_0)=\pm \int^1_0 \frac{2r^2_0{\mathrm{d}z}}{\bar{\alpha}\sqrt{G(z,r_0)}},
\end{equation} where
\begin{equation}\label{15}
G(z,r_0)= \Bigg(r^2_0 +(1-z)^2a^2\Bigg)^3\left[\frac{1}{\beta^2}-\frac{(1-z)^2\left(\sqrt{r^2_0 +(1-z)^2a^2}-2M(1-z)\right)}{ \Bigg(r^2_0 +(1-z)^2a^2\Bigg)^{3/2}}\right].
\end{equation}

By performing the series expansion of the function $G(z,r_0)$ at the point $z=0$, which corresponds to the limit $r\to {r_0}$, {and considering $\beta=\beta(r_0)$ given by Eq. (\ref{8})}, we obtain that:
\begin{equation}\label{16}
G(z,r_0) \simeq \Lambda_1(r_0)z + \Lambda_2(r_0)z^2,
\end{equation} where the expansion parameters $\Lambda_1$ and $\Lambda_2$ are defined as
\begin{equation}\label{17}
\Lambda_1(r_0)=2r^2_0\Bigg(r^2_0+a^2-{3}M\sqrt{r^2_0+a^2}\Bigg), \qquad \Lambda_1(r_0\to{r_m})=0, 
\end{equation}

\begin{equation}\label{18}
\Lambda_2(r_0)=\frac{r^2_0}{\sqrt{r^2_0+a^2}}\Bigg[3Mr^2_0 +\left(9a^2+r^2_0\right)\left(3M-\sqrt{r^2_0+a^2}\right)\Bigg], \quad
\Lambda_2(r_0\to{r_m})= (9M^2-a^2)^2.
\end{equation}
{While the expansion in Eq. (\ref{16}) is a general geometric feature near any turning point $r_0$, the actual strong-field limit regime relies on the specific initial condition where the impact parameter approaches its critical value ($\beta \to \beta_c$). This forces the turning point to be closely localized near the photon sphere ($r_0 \to r_m$), causing $\Lambda_1$ to vanish.}

Regarding the expansion coefficients of Eqs. (\ref{17}) and (\ref{18}) in the strong-field regime $r\to {r_m}$, when considering the integration expression of Eq. (\ref{14}), we have a logarithmic divergence and, therefore, it is necessary to separate this integral into two parts: a regular part $\Delta\phi_R(r_0)$ and a divergent part $\Delta\phi_D(r_0)$. Therefore, we have:
\begin{equation}\label{19}
\Delta\phi(r_0) = \Delta\phi_R(r_0) + \Delta\phi_D(r_0).
\end{equation}

Thus, the divergent part is defined by
{
\begin{eqnarray}\label{20}
\Delta\phi_D(r_0)&=& \frac{1}{\bar{\alpha}}\int^1_0\frac{2r^2_0 \mathrm{d}z}{\sqrt{\Lambda_1(r_0)z+\Lambda_2(r_0)z^2}}= 
-\frac{4r^2_0}{\bar{\alpha}\sqrt{\Lambda_2(r_0)}}\log\left(\sqrt{\Lambda_1(r_0)}\right) \nonumber \\
&+& \frac{4r^2_0}{\bar{\alpha}\sqrt{\Lambda_2(r_0)}}\log\left(\sqrt{\Lambda_2(r_0)}\,+\sqrt{\Lambda_1(r_0)+\Lambda_2(r_0)}\right),\qquad \mbox{when} \qquad r_0\to{r_m} \nonumber \\ 
\Delta\phi_D(r_m)&=& -\frac{2r^2_m} {\bar{\alpha}\sqrt{\Lambda_2(r_m)}}\log\left(\Lambda_1(r_m)\right) +\frac{2r^2_m}{\bar{\alpha}\sqrt{\Lambda_2(r_m)}}\log\left(4\Lambda_2(r_m)\right).\nonumber \\ 
\end{eqnarray}
}

In order to have some control over the divergent part of the integration, we need to expand the coefficient $\Lambda_1(r_0)$ and the impact parameter $\beta(r_0)$ over the radius of the photon sphere $r_m$. Thus, Eqs. (\ref{6C}) and (\ref{11}) become:
\begin{equation}\label{21}
    \beta\left(r_0\right) \simeq 3\sqrt{3}M + \frac{\sqrt{3}}{2M}\Bigg(\sqrt{r^2_0+a^2}-3M\Bigg)^2,
\end{equation}
\begin{equation}\label{22}
\Lambda_1(r_0) \simeq 2r^2_0\sqrt{r^2_0+a^2}\Bigg[\sqrt{6}M\left(\frac{\beta}{\sqrt{27M^2}}-1\right)^{1/2}\Bigg].
\end{equation}

Thus, substituting Eqs. (\ref{18}), (\ref{21}), and (\ref{22}) in the expression referring to the divergent part, Eq. (\ref{17}), we see that
\begin{equation}\label{23}
\Delta\phi_D= -\frac{1}{\bar{\alpha}}\log\Bigg(\frac{\beta}{\sqrt{27M^2}}-1\Bigg)+ \frac{2}{\bar{\alpha}}\log\left[\frac{2(9M^2-a^2)}{\sqrt{6}M^2\Bigg(3+\sqrt{6}\Bigg(\frac{\beta}{\sqrt{27M^2}}-1\Bigg)^{1/2}\Bigg)}\right].
\end{equation}

The equation above represents the analytical expression for the divergent part of the light deflection in the spacetime of a WH formed via a defect. Regarding the contribution related to the regular part of the gravitational deflection, we must consider the limit where the impact parameter, Eq. (\ref{6B}), assumes its critical value $r_0\to{r_m}$ and then use Eqs. (\ref{18}) and (\ref{15}) in that limit. Thus, the general expression for the regular part becomes
\begin{equation}\label{24}
\Delta\phi_R=\frac{1}{\bar{\alpha}} \Bigg(\int^{1}_0 \frac{2r^2_{m}}{\sqrt{G(z,r_{m})}}\mathrm{d}z 
-\int^{1}_{0}\frac{2r^2_{m}}{\sqrt{\Lambda_2(r_{m})}}\frac{\mathrm{d}z}{z}\Bigg).
\end{equation}

Therefore, the total deviation of light due to the presence of the gravitational field in this WH embedded in the global monopole spacetime is defined as $\delta\phi=\Delta\phi_D + \Delta\phi_R -\pi$. Since the expression for the regular part of the integration does not have an analytical solution, we will perform the numerical treatment.


\begin{figure}
    \centering
    \includegraphics[scale=0.41]{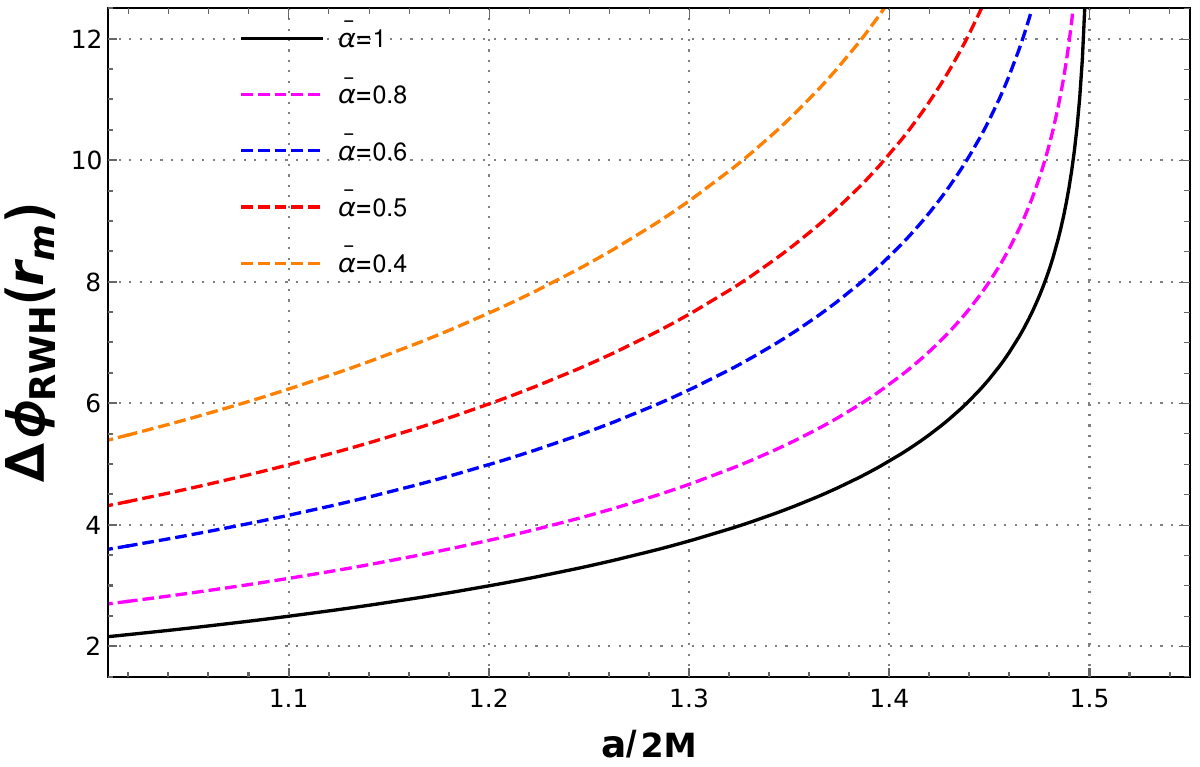}
     \includegraphics[scale=0.41]{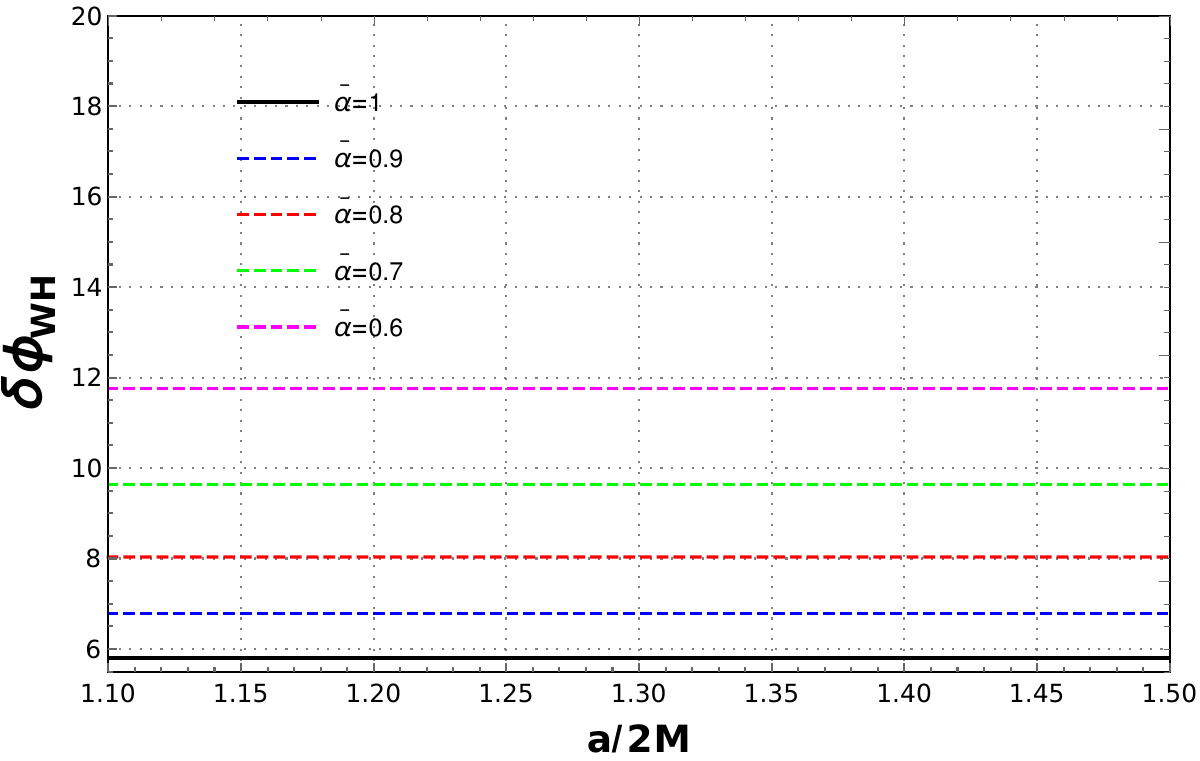}
    \caption{In (a) and (b) we have respectively the regular part of the integration (\ref{24}) and the angular deviation as a function of $a/2M$ for $\beta/2M=\frac{3\sqrt{3}}{2}+0.005$ and then we vary the charge on the global monopole.}
    \label{FORTE1}
\end{figure}

\begin{figure}
    \centering
    \includegraphics[scale=0.41]{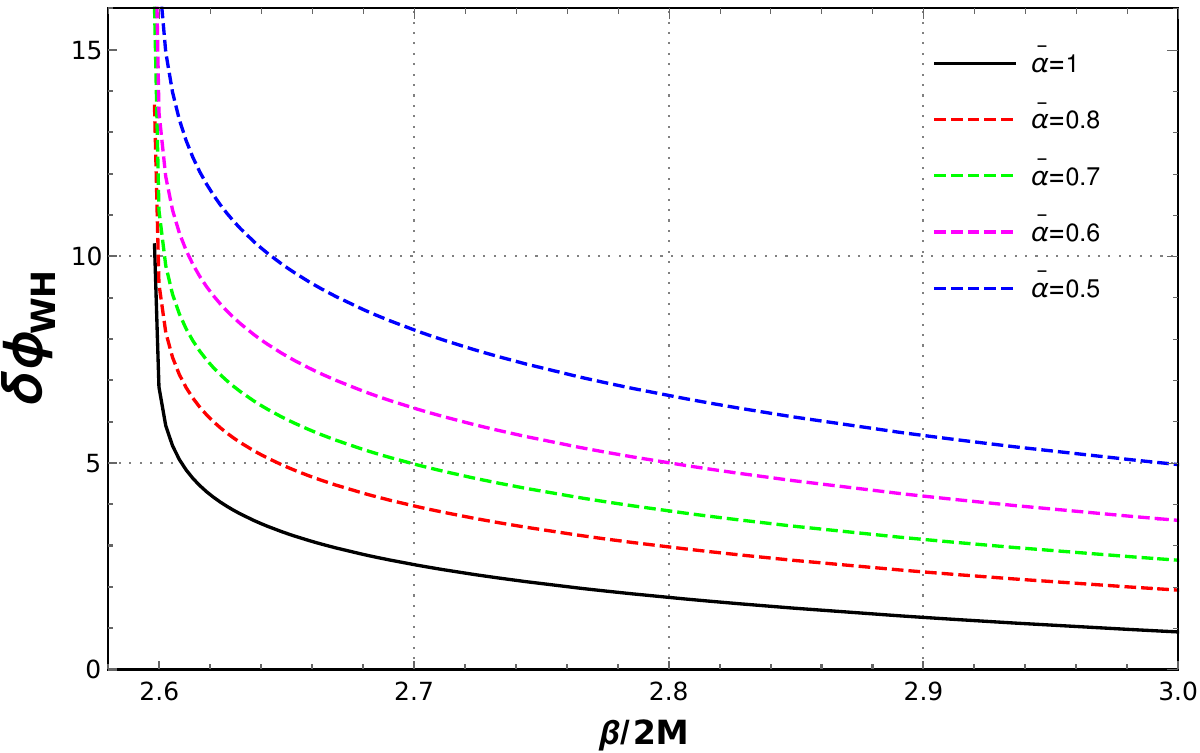}
      \includegraphics[scale=0.41]{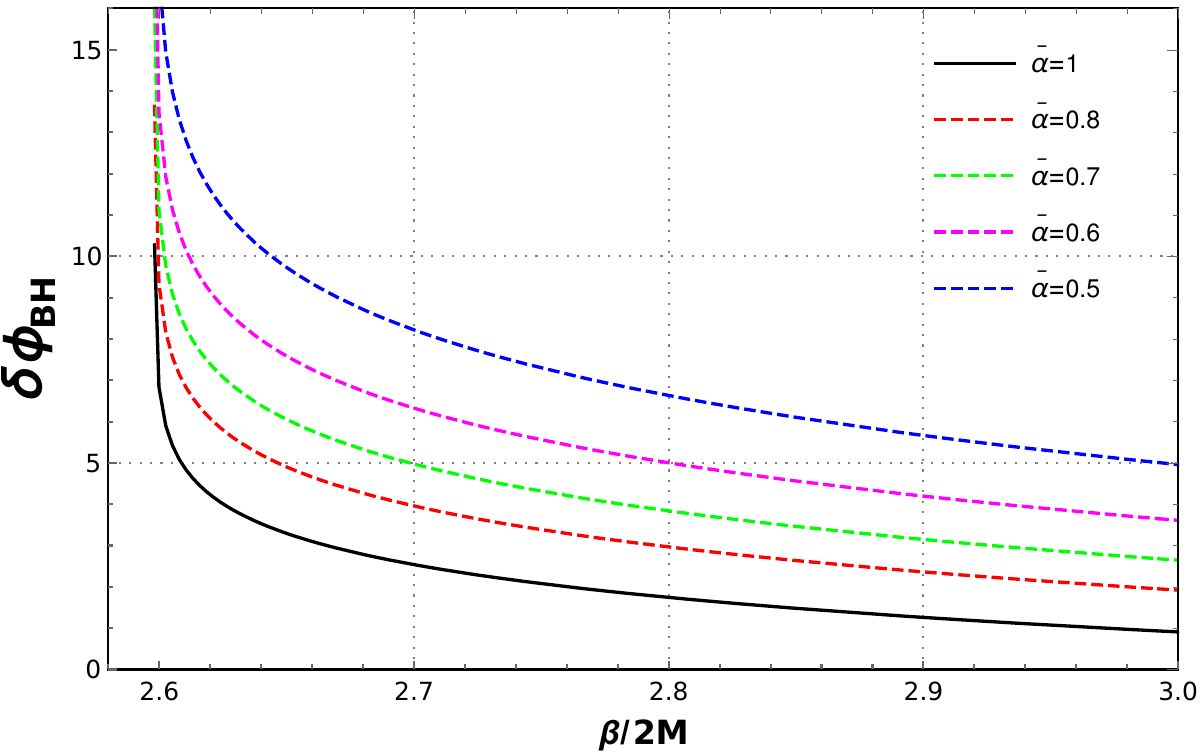}
    \caption{In (a) and (b) we have the angular deviation as a function of the impact parameter ratio $\beta/2M$ respectively for the WH and BH. In the case of the WH, we fix the ratio $a/2M=1.1$ and then vary the charge of the global monopole.}
    \label{FORTE2}
\end{figure}

In Fig. \ref{FORTE1}, in general terms, the first panel presents the graphical representation of the regular contribution of the integral (\ref{24}), while the second panel displays the angular deflection associated with the WH. In both cases, the analyzed quantities are shown as functions of the ratio $a/2M$. {To this end, several values for the global monopole charge were considered to examine the behavior of the curves. It is important to emphasize that, although these values are very close to unity for a typical unification scale \cite{B9K,B9A}, we also adopted more distinct parameters to clearly demonstrate their impact on the observables.} Thus, in both figures, the solid black curve represents the regular part of the integration (\ref{24}) and the angular deflection when the global monopole charge is turned off, whereas the dotted curves illustrate how the monopole charge modifies the solution. Consequently, the smaller the charge values, the more the curves tend to deviate from the baseline solution (black curve). Note that the angular deviation (figure on the right) for WH represents only a constant value when the global monopole charge is inactive (black curve). This is consistent with the Schwarzschild BH results \cite{ANEL2,ANEL3}. The variation in the numerical values of the global monopole charge represents only a shift in the curve associated with the BH, which corroborates the angular deviation in the {weak}-field regime (\ref{13}) that did not depend on the parameter associated with the throat $a$ of the WH.

In Fig. \ref{FORTE2}, we present the angular deflection of the WH and the BH as a function of the ratio between the impact parameter and the mass, $\beta/2M$. Similarly to what was discussed above, the solid black curve represents the case in which the global monopole charge is turned off. For the panel corresponding to the WH (figure on the left), special care is taken to fix the parameter $a/2M$, which characterizes the scenario of a traversable WH, while the values of the global monopole charge are varied. We observe that as the allowed values of the charge decrease, the curves tend to deviate increasingly from the reference solution (black curve). On the other hand, in the figure on the right, we perform the same analysis considering the limit in which the radius of the BH's throat tends to zero, $a \to 0$, recovering the results corresponding to the Schwarzschild BH. Thus, we examine how the monopole charge modifies the reference solution (black curve), which corresponds to the case in which the effects of the monopole are still absent, $\bar{\alpha} = 1$.


\section{Lens equation and observables}\label{sec4}
In this section, we will study the connection between the deflection of light in the weak-field regime, Eq. (\ref{13}), and in the strong-field regime, Eqs. (\ref{23}) and (\ref{24}), with $\delta\phi= \Delta\phi_R + \Delta\phi_D -\pi$, using the master equations of gravitational lensing in spacetime. Thus, we can construct physical quantities that can be theoretically observed. We consider lenses in the strong-field regime as a starting point.

\begin{figure}[htb!]
\centering
	{\includegraphics[width=0.75\textwidth]{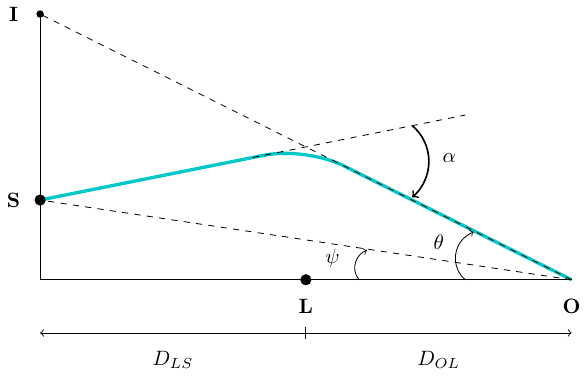}}
    \caption{Light angular deflection diagram.}
\label{LENS1}
\end{figure}

In Fig. (\ref{LENS1}), the panel visually illustrates the lens diagram. The light beam emitted by the source \textbf{S} is deflected by the presence of the WH located at \textbf{L}, subsequently propagating toward the observer \textbf{O}. The angular deflection of the light path is denoted by $\alpha$. The angular positions of the source and the image with respect to the optical axis, $\overline{LO}$, are represented by $\psi$ and $\theta$, respectively. Thus, considering that the source \textbf{S} is practically aligned with the lens \textbf{L}, a position in which the relativistic images should be more expressive \cite{B10,B11}, we then have the lens equation that relates the angular positions $\psi$ and $\theta$ defined as $\psi = \theta -\frac{D_{LS}}{D_{OS}}\Delta\alpha_n$. where $\Delta\alpha_n$ is the deflection angle subtracted from all the loops made by the photons before reaching the observer, that is, $\Delta\alpha_n = \alpha -2n\pi$. In this approach, we use the following approximation for the impact parameter $\beta\approx\theta{D_{OL}}$. 

{Then, the contributions to the angular deflection, Eqs. (\ref{23}) and (\ref{24}), are rewritten as $\alpha(\theta)= -\bar{a}\log\left(\frac{\theta{D_{OL}}}{\beta_c}-1\right) +\bar{b},$ where the critical impact parameter $\beta_c$ and the parameters $\bar{a}$, $\bar{b}$ are given as:}
\begin{eqnarray}\label{LT3}
\beta_c &=& 3\sqrt{3}M, \qquad \qquad \bar{a}= \frac{1}{\bar{\alpha}}, \\\label{LT4}
\bar{b}&=& \frac{2}{\bar{\alpha}}\log\left[\frac{2(9M^2-a^2)}{\sqrt{6}M^2\left[3+\sqrt{6}\left(\frac{\beta}{\beta_c}-1\right)^{1/2}\right]}\right] + \frac{(\Delta\phi_R -\bar{\alpha}\pi)}{\bar{\alpha}},
\end{eqnarray}
given that the regular part is given by Eq. (\ref{24}). To obtain $\Delta\alpha_n$, we expand $\alpha(\theta)$ close to $\theta = \theta^{0}_n$, where $\alpha(\theta^{0}_n)
 = 2n\pi$, and find that
 \begin{equation}\label{LT5}
\Delta\alpha_n= \frac{\partial\alpha}{\partial\theta} \Big|_{\theta=\theta^{0}_n}\left(\theta -\theta^{0}_n\right).
\end{equation}

{Considering the contribution to the angular deflection defined above as $\alpha(\theta)$ and given that $\theta=\theta^{0}_n$, we have}
\begin{equation}\label{LT6}
\theta^{0}_n= \frac{\beta_c}{D_{OL}}(1+e_n), \qquad \mbox{with} \quad e_n= e^{\frac{\bar{b}-2n\pi}{\bar{a}}}.
\end{equation}

{Replacing the contribution of angular deflection $\alpha(\theta)$} and (\ref{LT6}) in Eq. (\ref{LT5}), we have
\begin{equation}\label{LT7}
\Delta\alpha_n= - \frac{\bar{a}D_{OL}}{\beta_c{e_n}}\left(\theta-\theta^{0}_n\right).
\end{equation}

{Therefore, substituting Eq. (\ref{LT7}) into the angular position equation, we have}
\begin{eqnarray}\label{LT8}
\theta_n \approx \theta^{0}_n + \left(\frac{{e_n}\beta_c}{\bar{a}}\right) \frac{D_{OS}(\psi-\theta^{0}_n)}{D_{OL}D_{LS}}.
\end{eqnarray}

Although light deflection preserves surface brightness, gravitational lensing modifies the solid angle under which the source is observed. Consequently, the total flux received from a lensed image is proportional to its magnification $\mu_n$, which is defined by $\mu_n=\Big| \frac{\psi}{\theta}\frac{\partial\psi}{\partial\theta}\mid_{\theta=\theta^{0}_n}\Big|^{-1}$. {Therefore, substituting the angular position equation Eq. (\ref{LT7}), we obtain}

\begin{equation}\label{LT9}
\mu_n= \frac{e_n(1+e_n)}{\psi\bar{a}}\left(\frac{\beta_c}{D_{OL}}\right)^2\frac{D_{OS}}{D_{LS}}.
\end{equation}

The magnification decreases very rapidly as $n$ increases, implying that the brightness of the first image, $\theta_1$, dominates over the subsequent ones. Moreover, the presence of the factor  $\left(\frac{\beta_c}{D_{OL}}\right)^2$ indicates that the magnification remains intrinsically small. It is also observed that, in the limit  $\psi\to{0}$, corresponding to the maximal alignment between the source, the lens, and the observer, the magnification diverges, thereby enhancing the likelihood of detecting relativistic images.

\subsection{Observables in the strong-field limit}\label{sec41}

In the preceding sections, we expressed the relativistic image positions, as well as their fluxes, in terms of the expansion parameters $\bar{a}$, $\bar{b}$, and $\beta_c$. We now undertake the inverse procedure, reconstructing the expansion coefficients by taking the observational data as our starting point. This allows us to investigate the properties of the object producing the gravitational lens and, subsequently, to compare the observational results with the corresponding theoretical predictions. The impact parameter may be written in terms of $\theta_\infty$ \cite{B8},
\begin{equation}\label{LT10}
\beta_c= D_{OL}\theta_\infty.
\end{equation}

We will follow Bozza \cite{B8} and assume that only the outermost image $\theta_1$ is resolved as a single image while the others are encapsulated in $\theta_\infty$. Thus, Bozza defined the following observables,
\begin{eqnarray}\label{LT11}
s&=&\theta_1-\theta_\infty=\theta_\infty{e^{\frac{\bar{b}-2\pi}{\bar{a}}}}, \\\label{LT12}
\tilde{r} &=& \frac{\mu_1}{\sum^\infty_{n=2}\mu_n}=e^{\frac{2\pi}{\bar{a}}},
\end{eqnarray} where the parameters $\bar{a}$ and $\bar{b}$ are defined in the expressions (\ref{LT3}) and (\ref{LT4}).

In the expressions above, $\textbf{s}$ denotes the angular separation, while $\tilde{r}$ quantifies the relation between the flux of the first image and that of the remaining ones. These expressions can be inverted to recover the expansion coefficients. For the analysis of the observables, we assume that the object under consideration has an estimated mass of $4.4\times10^{6}M_{\odot}$ and is located at an approximate distance of $D_{OL}=8.5\,\text{kpc}$, values comparable to those associated with the BH at the center of our galaxy \cite{B12}. Knowing that the critical impact parameter, $\beta_c=3\sqrt{3}M$, does not depend on the radius of the throat, $a$, we then obtain it directly. In geometric units, we have the rescaling of the mass to $M\to \frac{MG}{c^2}$ and $\theta_\infty=26.5473\,\mu{arcsecs}$, which is the same used for the Schwarzschild BH.

\begin{figure}
    \centering
    \includegraphics[scale=0.41]{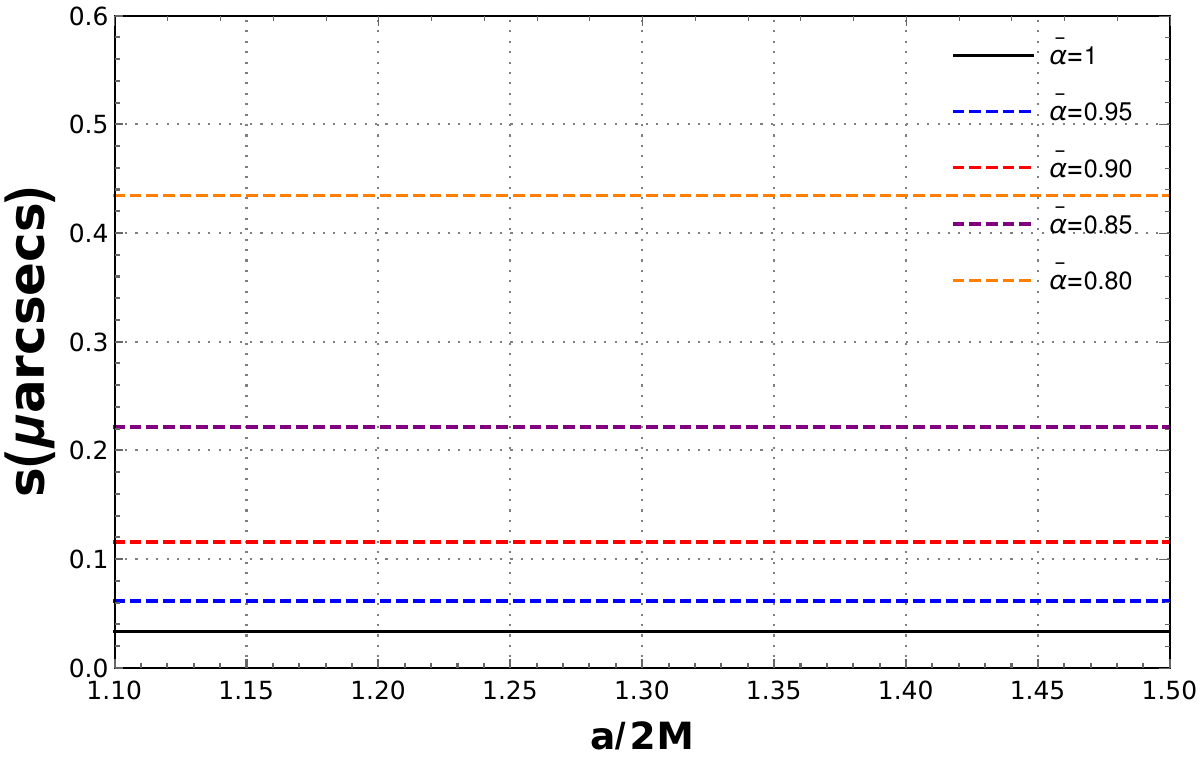}
      \includegraphics[scale=0.41]{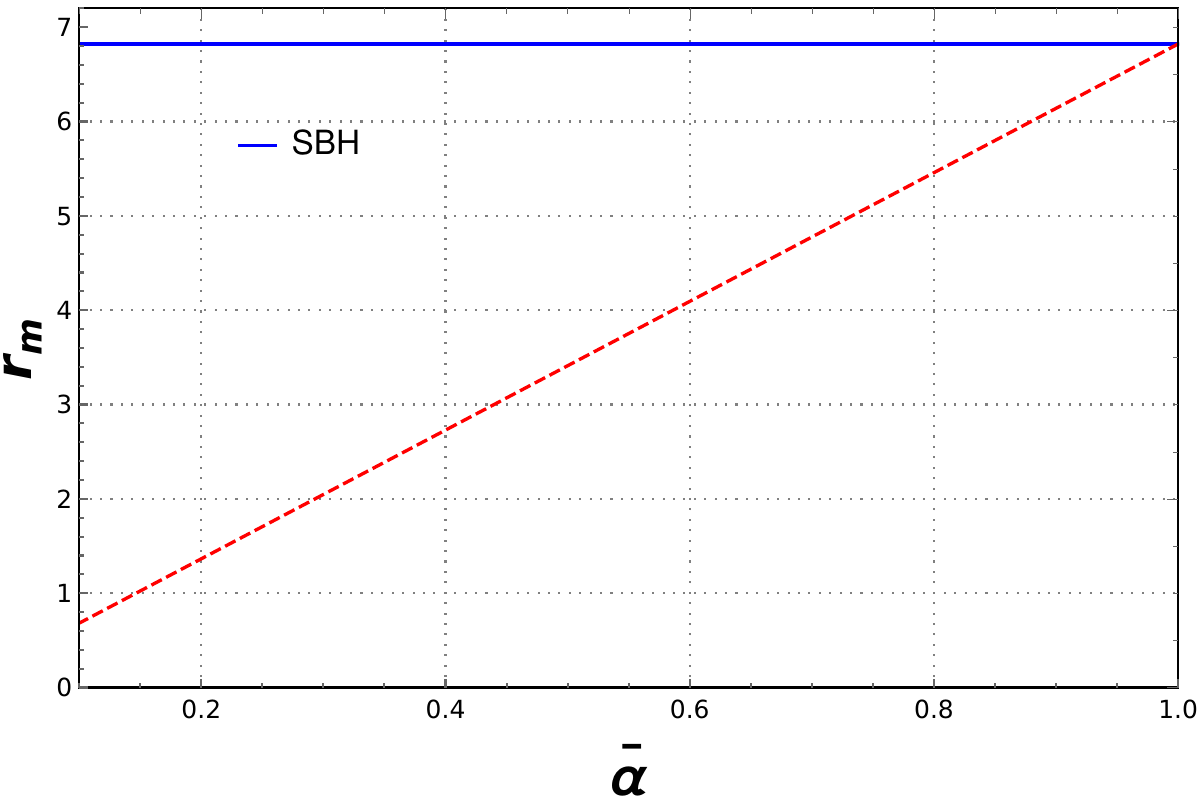}
    \caption{In (a) we have the angular separation $\textbf{s}$ as a function of $a/2M$ and varying the charge of the global monopole and in (b) $2.5\log_{10}\tilde{r}$ as a function of the charge of the monopole $\bar{\alpha}$.}
    \label{OBSERVAVEIS1}
\end{figure}

In Fig. (\ref{OBSERVAVEIS1}), we have a graphical representation of the angular separation $\textbf{s}$ and $2.5\log_{10}\tilde{r}$. For the angular separation varying as a function of the ratio between the parameters $a/2M$ considering some values of the global monopole charge and $2.5\log_{10}\tilde{r}$ also varying as a function of the global monopole charge. In the figure on the left, we have the variation of the angular separation as a function of the ratio $a/2M$ where the continuous black curve is fixed for traversable WH $a/2M=1.1$ where the charge of the global monopole is inactive. However, the dashed curves represent the variation of the monopole charge values. In the figure on the right, we have $2.5\log_{10}\tilde{r}$ varying as a function of the global monopole charge. The solid blue curve refers to the Schwarzschild BH, while the dashed red curve shows that when the global monopole charge $\bar{\alpha}\to{1}$ returns to the BH curve, this quantity does not depend on the parameter associated with the throat of the WH Eq. (\ref{LT12}). {Therefore, to construct table \ref{TAB1}, we used the data associated with the Schwarzschild BH located at the center of our galaxy, as mentioned in the previous paragraphs. Furthermore, we observed how the angular separation and the parameter that controls the connection between the flow of the first image and the others referring to the Schwarzschild BH are modified due to the presence of the global monopole.}

\begin{table}[!ht]
\centering
\caption{Observable quantities for a Schwarzschild BH as functions of the global monopole parameter $\bar{\alpha}$.}
\label{Tab:obs}

\renewcommand{\arraystretch}{1}

\begin{tabular}{c|c|c}
\hline\hline
$\bar{\alpha}$ 
& $s\;(\mu\mathrm{arcsecs})$ 
& $r_m\;(\mathrm{magnitudes})$ \\ 
\hline\hline
1.00 & 0.0332 & 6.8218 \\ \hline
0.95 & 0.0615 & 6.4808 \\ \hline
0.90 & 0.1156 & 6.1397 \\ \hline
0.85 & 0.2216 & 5.7986 \\ \hline
0.80 & 0.4342 & 5.4575 \\ \hline
0.75 & 0.8741 & 5.1164 \\ \hline
0.70 & 1.8181 & 4.7753 \\ \hline
0.65 & 3.9366 & 4.4342 \\ \hline\hline
\end{tabular}\label{TAB1}
\end{table}

{As shown in Table \ref{TAB1}, as the values of $\bar{\alpha}$ decrease, indicating a stronger influence of the monopole, the angular separation increases, leading to a wider gap between the first relativistic image and the subsequent ones. From $\bar{\alpha}=1$ to $\bar{\alpha}=0.65$, the angular separation increases by approximately two orders of magnitude, an improvement in the resolution of potential detections. Conversely, the decrease in magnitude also indicates a reduction in the brightness of the first image relative to the others. While the detectability of these observables remains a challenge for the current resolution of the Event Horizon Telescope, these signatures are prime candidates for future space-based interferometry. Given that the next-generation Event Horizon Telescope (ngEHT) project is expected to achieve resolutions below $\sim 20 \, \mu\text{arcsecs}$, these data could be instrumental in distinguishing between scenarios with and without the presence of a monopole charge.}

\begin{figure}
    \centering
    \includegraphics[scale=0.66]{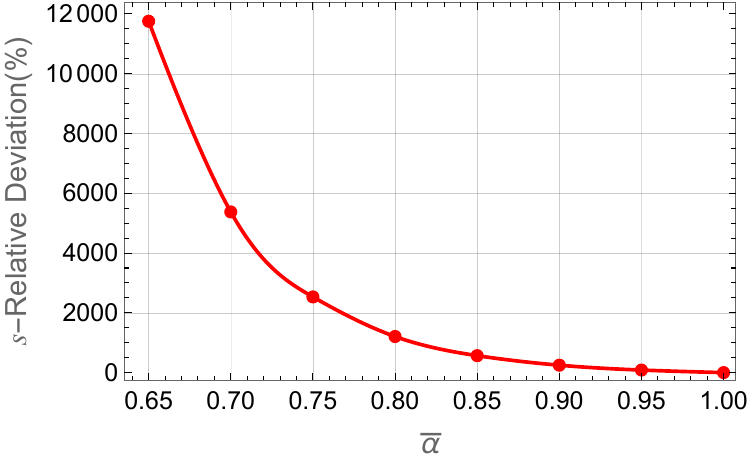}
      \includegraphics[scale=0.66]{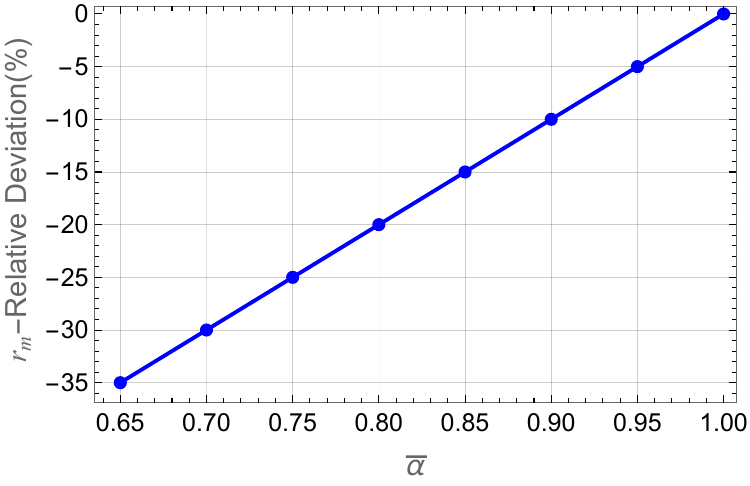}
    \caption{The left panel shows the percentage deviation of $s$ relative to the Schwarzschild case for the tabulated values of $\bar{\alpha}$, while the right panel displays the percentage deviation of $r_{m}$. }
    \label{desvio perc}
\end{figure}

{To demonstrate how the monopole charge causes a deviation in the observables that grows rapidly relative to the Schwarzschild case ($\bar{\alpha}=1$), we plotted in Fig. \ref{desvio perc} the percentage deviation of the observables with respect to the Schwarzschild values, $s_{Sch} = 0.0332 \, \mu\text{as}$ and $r_{m \ Sch} = 6.8218$ magnitudes.}

{The quantitative analysis of the relative deviation in the angular separation, presented in Fig.~\ref{desvio perc}, reveals the dramatic impact of the topological charge on the optical signatures of the WH. It is observed that, while for values of $\bar{\alpha}$ close to unity the deviation relative to the Schwarzschild case remains moderate (approximately $85\%$ for $\bar{\alpha}=0.95$), it undergoes an exponential growth as the influence of the monopole increases. At the lower bound of our analysis, $\bar{\alpha}=0.65$, the relative deviation reaches a magnitude of $11757\%$, corresponding to an increase of two orders of magnitude in the separation between the first and the remaining relativistic images. This behavior highlights a unique observational signature that is severely distinct from the usual vacuum solutions, providing a robust numerical criterion for distinguishing defect WH models in future interferometric experiments. Unlike the angular separation, the presence of the global monopole charge induces a systematic negative deviation in the luminous flux. For the scenario with the strongest topological influence ($\bar{\alpha}=0.65$), a decrease of approximately $35\%$ in the magnitude is observed in comparison with the Schwarzschild case. Physically, this result implies that the global monopole not only increases the separation between the relativistic images, but also consistently reduces the brightness of the primary image relative to the subsequent ones. The correlation between the extreme enhancement of the angular separation and the proportional reduction in magnitude constitutes a set of theoretical evidences that facilitate the parametrization of the model and reinforce its detectability by instruments with high photometric sensitivity and angular resolution, such as the ngEHT.}


\subsection{Observables in the weak-field limit}\label{sec42}

The next step is to analyze the observables in the weak-field regime, which means that the impact parameter is very large, $\beta\gg{M}$, so the light beam does not form loops. Therefore, the expression Eq. (\ref{13}) becomes
\begin{equation}\label{LTW1}
\delta\phi \sim  \frac{1}{\bar{\alpha}}\Bigg[\pi(1-\bar{\alpha})+\frac{4M}{\beta}\Bigg].
\end{equation}

Starting from the equation that relates the angular position of the source to that of the image, we shall consider the case of perfect alignment between the source, the compact object, and the observer, which corresponds to $\psi = 0$. In this way, we have to
\begin{equation}\label{LTW2}
\theta= \frac{D_{LS}}{D_{OS}}\Delta\alpha_n,
\end{equation} where $\Delta\alpha_n$ is given by Eq. (\ref{LTW1}). Therefore, substituting Eq. (\ref{LTW1}) into Eq. (\ref{LTW2}) and keeping only terms up to first order in the mass $M$, we obtain a second-degree algebraic equation, which when solved gives us the angular position $\theta_E$ for the Einstein ring:
\begin{equation}\label{LTW3}
\theta=\theta_E= \Bigg[\frac{D_{LS}\pi(1-\bar{\alpha})}{2\bar{\alpha}D_{OS}}\Bigg]\pm\Bigg(\frac{1}{2}\Bigg)\Bigg[\Bigg(\frac{D_{LS}\pi(1-\bar{\alpha})}{\bar{\alpha}D_{OS}}\Bigg)^2+\frac{16MD_{LS}}{\bar{\alpha}D_{OS}D_{OL}}\Bigg]^{1/2}.
\end{equation}
Therefore, it is clear from the above expression that the presence of the throat radius of the WH does not modify the position of the Einstein ring, depending only on the mass $M$ and the parameter associated with the charge of the global monopole $\bar{\alpha}$. From the expression above regarding the angular position of the Einstein ring, we can recover the expression referring to the Schwarzschild BH when the effects of the global monopole are {switched off}, $\bar{\alpha}\to{1}$. 
Thus, considering only the positive sign for the Einstein ring, we obtain the same expression (50) as in Ref. \cite{ANEL2} when the loop-quantum gravity (LQG) parameter tends to zero $a\to{0}$,
\begin{equation}\label{LTW31}
\theta=\theta_E= \sqrt{\frac{4MD_{LS}}{D_{OS}D_{OL}}}. 
\end{equation}

We can also calculate the radius of the Einstein ring from the approximation $\beta\approx\theta{D_{OL}}$ and using Eq. (\ref{LTW3}) \cite{ANEL1,ANEL2}. Therefore, we have
\begin{equation}\label{LTW4}
R_E= D_{OL}\theta_E= \Bigg[\frac{D_{OL}D_{LS}\pi(1-\bar{\alpha})}{2\bar{\alpha}D_{OS}}\Bigg]\pm\Bigg(\frac{1}{2}\Bigg)\Bigg[\Bigg(\frac{D_{LS}D_{OL}\pi(1-\bar{\alpha})}{\bar{\alpha}D_{OS}}\Bigg)^2+\frac{16MD_{LS}D_{OL}}{\bar{\alpha}D_{OS}}\Bigg]^{1/2}.
\end{equation} In this way, we can calculate the observables $R_E$ and $\theta_E$ taking into account some values for the parameter associated with the charge of the global monopole $\bar{\alpha}$ and the mass $M$. We will consider for the procedure of our analysis in this weak-field regime the lensing for a bulge star \cite{BOJO}. To this end, we consider the following values for the parameters $D_{OL}=4\,kpc$ and $D_{OS}=8\,kpc$. {Thus, in table \ref{TAB2}, we assign some possible values to the charge of the global monopole and then observe how the radius of the Einstein ring and the angular position respond to this variation.} It is worth noting that these results do not depend on the throat radius of the WH and therefore we can say that it is equivalent to Schwarzschild BH, being modified due to the charge of the global monopole.

\begin{table}[!ht]
\centering
\caption{Observable quantities as functions of the global monopole parameter $\bar{\alpha}$.}
\label{Tab:obs2}

\renewcommand{\arraystretch}{1}

\begin{tabular}{c|c|c}
\hline\hline
$\bar{\alpha}$ 
& $R_E\;(\mathrm{km})$ 
& $\theta_E\;(\mathrm{arcsecs})$ \\ 
\hline\hline
1.00 & $1.27\times10^{12}$ & 2.1200 \\ \hline
0.99 & $1.96\times10^{15}$ & 3272.7 \\ \hline
0.98 & $3.96\times10^{15}$ & 6612.3 \\ \hline
0.97 & $6.00\times10^{15}$ & 10020.6 \\ \hline
0.96 & $8.08\times10^{15}$  & 13500.0 \\ \hline
0.95 & $10.20\times10^{15}$ & 17052.6 \\ \hline
0.94 & $12.38\times10^{15}$  & 20680.9 \\ \hline\hline
\end{tabular} \label{TAB2}
\end{table}


\section{\label{MQ}Quasinormal modes}    

{\subsection{WKB tecnique}}

This section examines the quasinormal mode spectrum associated with the spacetime under consideration. By analyzing linear perturbations around the background geometry, we characterize the complex frequencies that govern the oscillatory response and decay of the system following a small disturbance. The dependence of these modes on the relevant parameters is explored, allowing us to assess how deviations from the reference configuration modify the dynamical relaxation toward equilibrium. It is worth mentioning that we shall focus on the scalar perturbations only. Initially, let us consider the most general form for a spherical symmetric metric before proceed further with the calculations
\begin{align}
\mathrm{d}s^{2} &= -A(r)\,\mathrm{d}t^{2} + B(r)\,\mathrm{d}r^{2} + C(r)\Big(\mathrm{d}\theta^{2}+\sin^{2}\theta\,\mathrm{d}\varphi^{2}\Big),
\qquad A(r)>0,\;B(r)>0,\;C(r)>0 .
\end{align}
A massless scalar field $\Phi$ obeys the Klein--Gordon equation
\begin{align}
\square \Phi
= \frac{1}{\sqrt{-g}}\partial_\mu\!\left(\sqrt{-g}\,g^{\mu\nu}\partial_\nu \Phi\right)=0,
\qquad
\sqrt{-g}=\sqrt{A(r)B(r)}\,C(r)\sin\theta .
\end{align}

To carry out the separation of variables and isolate the effective potential governing scalar perturbations, we decompose the field as follows:
\begin{align}
\Phi(t,r,\theta,\varphi)
= e^{-i\omega t}\,Y_{l m}(\theta,\varphi)\,R_{l\omega}(r),
\end{align}
where the spherical harmonics satisfy
\begin{align}
\Delta_{S^{2}}Y_{l m}=-l(l+1)Y_{l m}.
\end{align}
The radial function then fulfills
\begin{align}
\frac{1}{\sqrt{A B}\,C}\frac{\mathrm{d}}{\mathrm{d}r}
\!\left(\sqrt{\frac{A}{B}}\,C\,\frac{\mathrm{d}R}{\mathrm{d}r}\right)
+\left(\frac{\omega^{2}}{A}-\frac{l(l+1)}{C}\right)R_{l \omega}=0 .
\end{align}
Here, by introducing the tortoise coordinate, which reads
\begin{align}
\frac{\mathrm{d}r^*}{\mathrm{d}r}=\sqrt{\frac{B(r)}{A(r)}},
\end{align}
where after substituting the corresponding terms $A(r)$ and $B(r)$, we get
\begin{equation}
  r^{*} =  \frac{ \sqrt{a^2+r^2}+2 M \ln \left(\sqrt{a^2+r^2}-2 M\right)}{\bar{\alpha} }.
\end{equation}

Furthermore, by redefining the radial function as \cite{Santos:2025xbk}
$ \psi_{l\omega}(r)=\sqrt{C(r)}\,R_{l\omega}(r)$, the radial equation becomes
\begin{align}
\frac{\mathrm{d}^{2}\psi}{\mathrm{d}r^{*2}}
+\Big(\omega^{2}-\mathrm{V}_{s}(r)\Big)u=0 ,
\end{align}
with the effective potential being
\begin{align}
\mathrm{V}_{s}(r)
= A(r)\,\frac{l(l+1)}{C(r)}
+ \frac{1}{\sqrt{C(r)}}\frac{\mathrm{d}^{2}}{\mathrm{d}r^{*2}}
\!\Big(\sqrt{C(r)}\Big).
\end{align}

Expressed purely in terms of $r$--derivatives (prime $=\frac{\mathrm{d}}{\mathrm{d}r}$), this reads \cite{Santos:2025xbk}
\begin{align}
\mathrm{V}_{s}(r)
= A\,\frac{l(l+1)}{C}
+ \frac{A}{B}\left[
\frac{1}{2}\frac{C''}{C}
-\frac{1}{4}\left(\frac{C'}{C}\right)^{2}
+\frac{1}{4}\left(\frac{A'}{A}-\frac{B'}{B}\right)\frac{C'}{C}
\right].
\end{align}
After some algebraic manipulations, we obtain 
\begin{equation}
\label{effevssdsd}
\mathrm{V}_{s}(r,a,\Bar{\alpha}) = f(r) \left(\frac{l(l+1) }{\Sigma (r)^2}+\frac{2 \bar{\alpha} ^2 M}{\Sigma (r)^3}\right).
\end{equation} 

{Since in the present analysis we consider the traversable WH regime, $a>2M$, the geometry has no event horizon and connects two asymptotically flat regions. Moreover, from Eq. (\ref{effevssdsd}), the scalar effective potential satisfies $V_s(r)\to 0$ as $r\to \pm\infty$. Therefore, in terms of the tortoise coordinate, the asymptotic solutions reduce to free plane waves at both ends of the spacetime. Assuming the time dependence $e^{-i\omega t}$, the quasinormal mode boundary conditions are taken to be purely outgoing at both asymptotic regions, namely
\begin{equation}
\psi \sim e^{+i\omega r_*}, \qquad r_* \to +\infty,
\end{equation}
and
\begin{equation}
\psi \sim e^{-i\omega r_*}, \qquad r_* \to -\infty.
\end{equation}
In this case, the effective potential forms a single barrier, so the sixth-order WKB quantization condition keeps the same formal structure as in the BH case, although the physical interpretation of the boundary conditions is different.}

\begin{figure}
    \centering
    \includegraphics[scale=0.53]{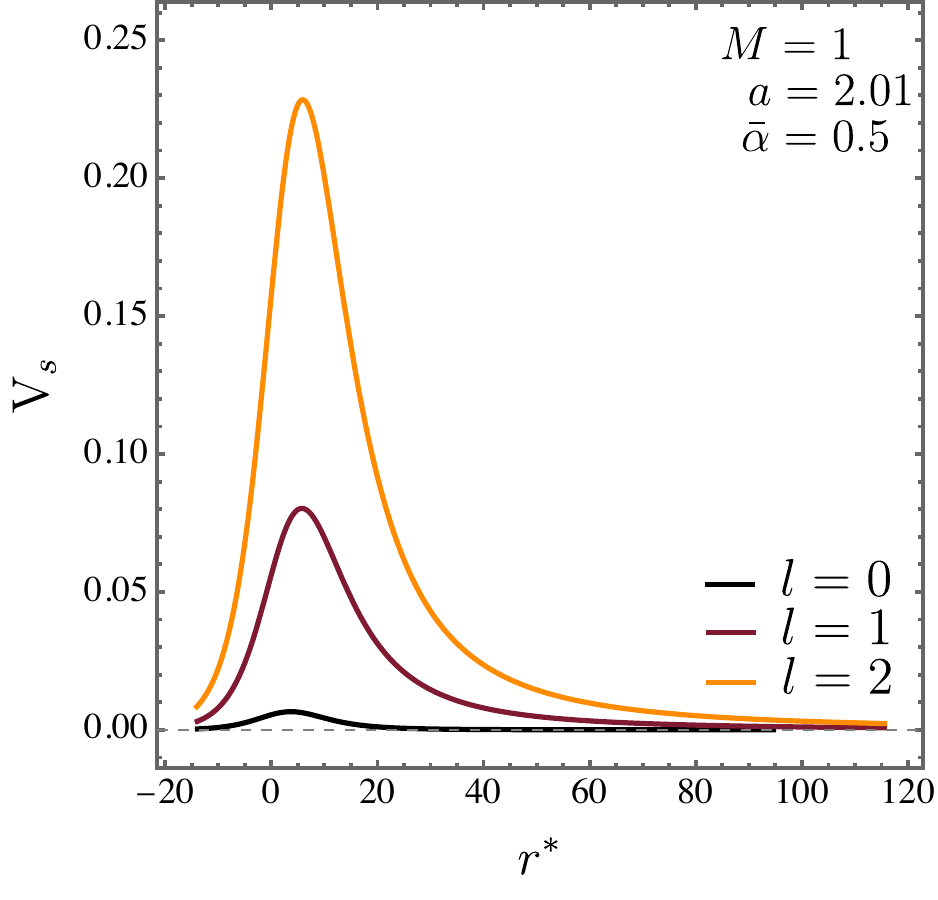}
    \caption{The effective potential governing scalar perturbations is shown as a function of the tortoise coordinate $r^{*}$ for different angular modes, namely $l=0$, $l=1$, and $l=2$, with the parameters fixed at $M=1$, $a=2.01$, and $\Bar{\alpha}=0.5$.}
    \label{vtorrr}
\end{figure}

In Fig. (\ref{vtorrr}), the effective potential $\mathrm{V}_{s}$ is displayed as a function of the tortoise coordinate. The plot shows that increasing the angular momentum parameter $l$ raises the height of the potential barrier. Another notable feature is that the barrier exhibits a smooth, sine--like shape. Under these conditions, the WKB approximation constitutes a suitable framework for computing the quasinormal modes, which we adopt in the following analysis.

The determination of the quasinormal frequencies proceeds from a local analysis centered on the extremum of the effective potential, where the wave dynamics undergo a transition between classically allowed and forbidden regions. In this framework, the potential peak sets the location of the turning points, around which the perturbation equation is treated perturbatively. By expressing the radial solution as a series expansion about this point and enforcing continuity through a WKB matching scheme, one arrives at approximate complex eigenvalues associated with the ringdown phase of the BH. Adopting the higher--order WKB formulation developed by Konoplya, these eigenfrequencies are constrained by \cite{Konoplya:2011qq}
\ie
\frac{i\bigl(\omega_{n}^{2}-\mathrm{V}_{0}\bigr)}{\sqrt{-2\mathrm{V}_{0}^{''}}}-\sum_{k=2}^{6}\Lambda_{k}=n+\frac{1}{2}.
\fe
Here, $\mathrm{V}_{0}^{''}$ denotes the curvature of the effective potential at its extremal point $r_{0}$. The quantities $\Lambda_{k}$ encode the hierarchy of higher--order contributions, each built from $\mathrm{V}_{0}$ and its successive derivatives, and their inclusion naturally refines the accuracy of the resulting quasinormal spectra. All numerical determinations of the quasinormal frequencies presented in what follows are obtained using the publicly available implementation described in Ref.~\cite{Konoplya:2019hlu}.

The numerical values reported in Tables~\ref{qnmssccaar1}–\ref{qnmssccaar2} correspond to the quasinormal spectra for multipole indices $l=1$, and $l=2$, evaluated for several combinations of the parameter $a$ and the deformation parameter $\Bar{\alpha}$, while the mass was kept fixed at $M=1$. A clear pattern emerged from these data. When $\Bar{\alpha}$ was held at $\Bar{\alpha}=0.7$ and the parameter $a$ was increased, the real components of the frequencies shifted downward, whereas the imaginary components decreased in magnitude (for $\omega_{0}$ and $\omega_{1}$). In contrast, fixing $a=2.01$ and varying $\Bar{\alpha}$ led to simultaneous growth of both the real and imaginary (its magnitude) parts (for $\omega_{0}$).

These trends translated directly into the dynamical response of the perturbations. Larger real parts signaled more rapid oscillatory behavior, while reduced imaginary parts corresponded to slower decay. As a result, configurations with higher values of $a$ supported longer--lived perturbations for a given $\Bar{\alpha}$ (for $\omega_{0}$ and $\omega_{1}$), indicating a tendency toward increased stability. The opposite tendency appeared when $\Bar{\alpha}$ was raised at fixed $a$, a behavior compatible with the previously discussed reduction of the effective potential barrier.

Regarding the overtone structure, the hierarchy followed the expected ordering: modes with higher overtone number, such as $\omega_{1}$ and $\omega_{2}$, decayed more quickly and oscillated at lower frequencies than the fundamental mode. In addition, the dominant contribution to the late--time signal was therefore governed by $\omega_{0}$, while higher overtones remained relevant only at earlier stages of the evolution.

\begin{table}[!h]
\begin{center}
\caption{\label{qnmssccaar1}The tabulated results present the scalar quasinormal frequencies $\omega_n$ associated with the monopole configuration ($l=1$), computed with the mass parameter set to $M=1$. The numerical values were evaluated by means of the sixth–order WKB method, allowing for detailed comparison through different choices of the parameters $a$ and $\Bar{\alpha}$.}
\begin{tabular}{c| c | c | c} 
 \hline\hline\hline 
 $M$ \quad \,\, $a$ \quad\quad  $\Bar{\alpha}$  & $\omega_{0}$ & $\omega_{1}$ & $\omega_{2}$  \\ [0.2ex] 
 \hline 
\,  1.0, \,  2.01, \,  0.7  & 0.277949 - 0.0684704$i$ & 0.246424 - 0.213125$i$  & 0.187649 - 0.382069$i$   \\
 
 \, 1.0, \,  2.05, \,  0.7  & 0.277877 - 0.0671686$i$ & 0.244834 - 0.210105$i$ & 0.183796 - 0.379362$i$ \\
 
\,  1.0, \,  2.10, \,  0.7  & 0.277723 - 0.0655075$i$ & 0.242188 - 0.206461$i$  & 0.172720 - 0.380584$i$   \\
 
 \, 1.0, \,  2.20, \,  0.7  & 0.277097 - 0.0620520$i$ & 0.222279 - 0.202719$i$ & 0.060586 - 0.519195$i$ \\
\hline\hline\hline 
 $M$ \quad \,\, $a$ \quad\quad  $\Bar{\alpha}$ & $\omega_{0}$ & $\omega_{1}$ & $\omega_{2}$  \\ [0.2ex] 
 \hline 
 \,  1.0, \,  2.05, \,  0.80  & 0.284616 - 0.0672762$i$ & 0.252136 - 0.209904$i$ &  0.191577 - 0.377756$i$  \\

  \,  1.0, \,  2.05, \,  0.85  & 0.288261 - 0.0673435$i$ & 0.256076 - 0.209853$i$ &  0.195806 - 0.377027$i$  \\
 
\,  1.0, \,  2.05, \,  0.90  & 0.292081 - 0.0674198$i$ & 0.260198 - 0.209840$i$  & 0.200253 - 0.376360$i$   \\
 
 \, 1.0, \,  2.05, \,  0.99  & 0.299380 - 0.0675795$i$ & 0.268053 - 0.209911$i$ & 0.208780 - 0.375335$i$ \\
   [0.2ex] 
 \hline \hline \hline 
\end{tabular}
\end{center}
\end{table}

\begin{table}[!h]
\begin{center}
\caption{\label{qnmssccaar2} The tabulated results present the scalar quasinormal frequencies $\omega_n$ associated with the monopole configuration ($l=2$), computed with the mass parameter set to $M=1$. The numerical values were evaluated by means of the sixth–order WKB method, allowing for detailed comparison through different choices of the parameters $a$ and $\Bar{\alpha}$.}
\begin{tabular}{c| c | c | c} 
 \hline\hline\hline 
 $M$ \quad \,\, $a$ \quad\quad  $\Bar{\alpha}$  & $\omega_{0}$ & $\omega_{1}$ & $\omega_{2}$  \\ [0.2ex] 
 \hline 
\,  1.0, \,  2.01, \,  0.7  & 0.474230 - 0.0700047$i$ & 0.454627 - 0.212585$i$  & 0.415788 - 0.363435$i$    \\
 
 \, 1.0, \,  2.05, \,  0.7  & 0.474181 - 0.0687802$i$ & 0.453604 - 0.209253$i$ & 0.413189 - 0.358815$i$  \\
 
\,  1.0, \,  2.10, \,  0.7  & 0.474079 - 0.0671994$i$ & 0.452120 - 0.205075$i$ & 0.409209 - 0.353209$i$   \\
 
 \, 1.0, \,  2.20, \,  0.7  & 0.473713 - 0.0638743$i$ & 0.447794 - 0.196562$i$ & 0.386990 - 0.344806$i$ \\
\hline\hline\hline 
 $M$ \quad \,\, $a$ \quad\quad  $\Bar{\alpha}$ & $\omega_{0}$ & $\omega_{1}$ & $\omega_{2}$  \\ [0.2ex] 
 \hline 
 \,  1.0, \,  2.05, \,  0.80  & 0.478097 - 0.0688304$i$ & 0.457650 - 0.209363$i$ & 0.417477 - 0.358860$i$  \\

 \,  1.0, \,  2.05, \,  0.85  & 0.480241 - 0.0688582$i$ & 0.459863 - 0.209426$i$ & 0.419823 - 0.358888$i$  \\
 
\,  1.0, \,  2.05, \,  0.90  & 0.482506 - 0.0688877$i$ & 0.462203 - 0.209492$i$  & 0.422301 - 0.358921$i$  \\
 
 \, 1.0, \,  2.05, \,  0.99  & 0.486889 - 0.0689453$i$ & 0.466726 - 0.209623$i$ & 0.42709 - 0.358992$i$ \\
   [0.2ex] 
 \hline \hline \hline 
\end{tabular}
\end{center}
\end{table}


{

\subsection{Eikonal correspondence between the photon sphere and quasinormal modes}

Before closing this section, it is useful to make the connection between the quasinormal spectrum and the critical null orbit more explicit. In the eikonal regime, $l \gg 1$, the scalar effective potential is dominated by the angular term, so that
\begin{equation}
V_{s}(r) \approx l(l+1)\frac{f(r)}{\Sigma^{2}(r)}.
\end{equation}
Therefore, the position of the maximum of the potential is determined by
\begin{equation}
\frac{\mathrm{d}}{\mathrm{d}r}\left(\frac{f(r)}{\Sigma^{2}(r)}\right)=0,
\end{equation}
which is precisely the same condition that selects the photon sphere. Hence, in the geometric--optics limit, the peak of the perturbative potential is governed by the same unstable null circular orbit discussed in the geodesic analysis.

Using the notation already introduced above, the orbital frequency $\Omega_{c}$ and the Lyapunov exponent $\lambda$ associated with the unstable photon sphere are given by
\begin{equation}
\Omega_{c}=\left.\sqrt{\frac{A(r)}{C(r)}}\right|_{r=r_{m}},
\qquad
\lambda=\left.\sqrt{\frac{A(r)C''(r)-A''(r)C(r)}{2B(r)C(r)}}\right|_{r=r_{m}},
\end{equation}
where, for the present WH geometry,
\begin{equation}
\nonumber
A(r)=f(r),
\qquad
B(r)=\frac{1}{g(r)f(r)},
\qquad
C(r)=\Sigma^{2}(r)=r^{2}+a^{2}.
\end{equation}
Evaluating these quantities at the photon sphere radius $r_{m}=\sqrt{9M^{2}-a^{2}}$, one has $\Sigma(r_{m})=3M$ and $f(r_{m})=1/3$. As a consequence,
\begin{equation}
\Omega_{c}
=
\left.\sqrt{\frac{f(r)}{\Sigma^{2}(r)}}\right|_{r=r_{m}}
=
\frac{1}{3\sqrt{3}M},
\end{equation}
while for the Lyapunov exponent one obtains
\begin{equation}
\lambda
=
\frac{\bar{\alpha}}{3\sqrt{3}M}.
\end{equation}
Thus, the eikonal quasinormal frequencies assume the standard asymptotic form
\begin{equation}
\omega_{l n}^{\rm eik}
\approx
l\,\Omega_{c}
-
i\left(n+\frac{1}{2}\right)\lambda
=
\frac{1}{3\sqrt{3}M}
\left[
l
-
i\bar{\alpha}\left(n+\frac{1}{2}\right)
\right].
\end{equation}

This result clarifies the role of the two parameters. At leading eikonal order, the throat parameter $a$ does not appear explicitly in the above expression. Although it shifts the coordinate location of the photon sphere, the combination $\Sigma(r_{m})$ remains equal to $3M$, and the corresponding cancellations remove the explicit dependence on $a$ from both $\Omega_{c}$ and $\lambda$. On the other hand, the monopole parameter $\bar{\alpha}$ does not modify the leading oscillation frequency, but it directly rescales the instability timescale of the null circular orbit and therefore changes the damping rate. In particular, smaller values of $\bar{\alpha}$ imply smaller $\lambda$, which corresponds to longer--lived eikonal modes.

For a representative configuration with $M=1$ and $\bar{\alpha}=0.7$, one finds
\begin{equation}
\Omega_{c}\simeq 0.19245,
\qquad
\lambda\simeq 0.13472,
\end{equation}
so that
\begin{equation}
\omega_{l n}^{\rm eik}
\approx
0.19245\,l
-
i\,0.13472\left(n+\frac{1}{2}\right).
\end{equation}
Since the numerical results presented in Tables IV and V correspond to the low--multipole sector, namely $l=1$ and $l=2$, an exact quantitative agreement with the eikonal formula is not expected. Even so, this asymptotic relation makes clear that the geometric origin of the large--$l$ spectrum is the same unstable photon sphere that controls the critical impact parameter and the shadow, whereas the residual dependence observed in the WKB data for low $l$ is associated with subleading corrections beyond the strict eikonal limit.

}


\section{\label{TDOMAIN}Time--domain solution}

Regarding the time evolution of scalar disturbances requires an approach that treats the dynamics directly, rather than relying solely on spectral information extracted in the frequency domain. A time--domain formulation tracks the propagation of the field itself and makes explicit how quasinormal ringing governs both the decay profile and the scattering response. Because the associated effective potentials typically exhibit nontrivial features, the numerical evolution must be handled with care; stability and precision can only be ensured through an appropriate computational strategy. To meet these requirements, we employ a characteristic evolution scheme, first formulated by Gundlach et. al.~\cite{Gundlach:1993tp}, which is well suited for this class of problems.

Following the numerical strategies implemented in Refs.~\cite{Bolokhov:2024ixe,Skvortsova:2024wly,Guo:2023nkd,Baruah:2023rhd,Shao:2023qlt,Lutfuoglu:2025kqp,Santos:2025xbk,Yang:2024rms}, the wave equation is rewritten in terms of double--null coordinates, defined by $u=t-r^{*}$ and $v=t+r^{*}$. This change of variables simplifies the causal structure of the evolution and renders the numerical integration more efficient. In these coordinates, the perturbation equation acquires a form that is particularly convenient for implementation, which we summarize below
\ie
\left(4 \frac{\partial^{2}}{\partial u \, \partial v} + V(u,v)\right) \Tilde{\psi} (u,v) = 0.
\fe

{Furthermore, the potential used in the time--domain integration is the scalar effective potential written in terms of the radial coordinate and then evaluated along the null grid through $r=r(u,v)$. More precisely, we write
\begin{equation}
V(u,v)\equiv V_s\bigl(r(u,v)\bigr)
=
f\bigl(r(u,v)\bigr)
\left[
\frac{l(l+1)}{\Sigma(r(u,v))^2}
+\frac{2\bar{\alpha}^{\,2}M}{\Sigma(r(u,v))^3}
\right],
\end{equation}
The function $r(u,v)$ is not obtained in closed analytical form; instead, it is determined numerically from the tortoise-coordinate relation
\begin{equation}
r^*=\frac{v-u}{2}.
\end{equation}}

The numerical implementation is carried out by replacing the continuous differential equation with a lattice representation defined on a discrete set of points. The computational domain is partitioned into a mesh, and the evolution equation is approximated by finite differences evaluated on this grid. Within this framework, the field at each node is determined from previously computed values, so that the waveform is advanced step by step along the grid, allowing its temporal development to be followed explicitly
\ie
\Tilde{\psi}(N) = -\Tilde{\psi}(S) + \Tilde{\psi}(W) + \Tilde{\psi}(E) - \frac{h^{2}}{8}V(S)\Big[\Tilde{\psi}(W) + \Tilde{\psi}(E)\Big] + \mathcal{O}(h^{4}).
\fe

The numerical evolution is organized by first laying out a regular lattice in the $(u,v)$ plane, characterized by a fixed increment $h$ along both null directions. The integration advances cell by cell across this lattice. For each cell, three vertices are already known, while the fourth—located at the future corner—must be reconstructed from the evolution scheme. The past point $(u,v)$ serves as the base of the cell, its neighbors along the null directions provide auxiliary input, and the value at the forward vertex is obtained through the update rule.

The procedure is initialized by prescribing data on two intersecting null hypersurfaces, specified by $u=u_{0}$ and $v=v_{0}$. These boundaries anchor the entire computation. On the incoming null line $u=u_{0}$, the scalar disturbance is seeded through a localized wave packet, chosen as a Gaussian distribution peaked at $v=v_{c}$ with width $\sigma$. Starting from this initial configuration, the algorithm propagates the field across the grid, generating the full time--domain evolution of the perturbation
\ie
\Tilde{\psi}(u = u_{0},v) = A e^{-(v-v_{0})^{2}}/2\sigma^{2}, \,\,\,\,\,\, \Tilde{\psi}(u,v_{0}) = \Tilde{\psi}_{0}.
\fe

The dynamical evolution is triggered by fixing boundary conditions along the null segment $v=v_{0}$, where the field is set to $\tilde{\psi}(u,v_{0})=0$. This choice provides a clean and numerically reasonable starting point. From this boundary, the algorithm advances the solution stepwise along increasing values of $v$ at fixed $u$, respecting the causal ordering dictated by the double--null discretization. To simplify the analysis and avoid unnecessary sources of instability, the study is restricted to massless scalar disturbances, while the BH mass is kept fixed at $M=1$ throughout. The initial excitation is introduced as a localized Gaussian pulse centered at $v=0$ with width $\sigma=1$. The computational domain in the $(u,v)$ plane spans the interval $[0,1000]$ in both directions and is sampled uniformly with step size $h=0.1$, which proves sufficient to resolve the oscillatory phase and the subsequent decay of the signal.

\begin{figure}
    \centering
    \includegraphics[scale=0.51]{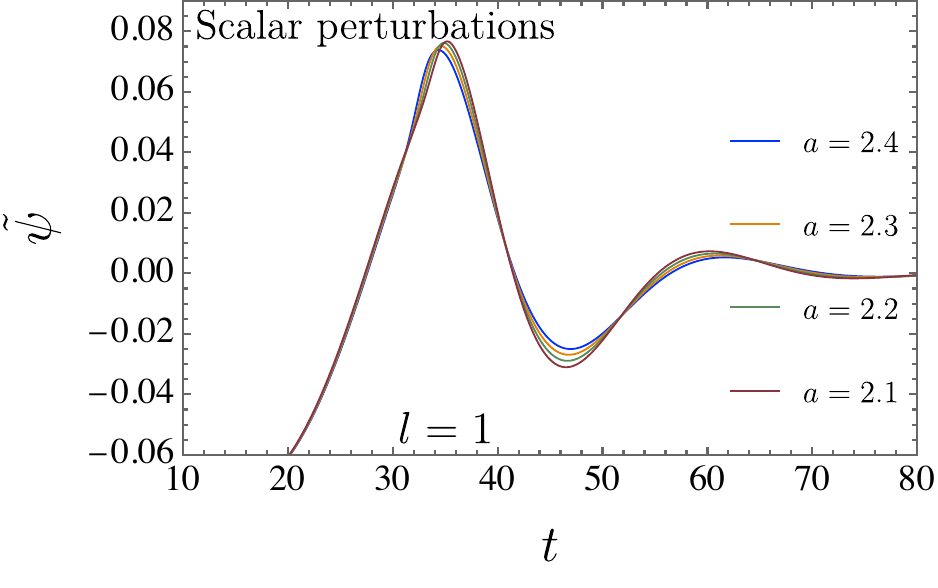}
     \includegraphics[scale=0.51]{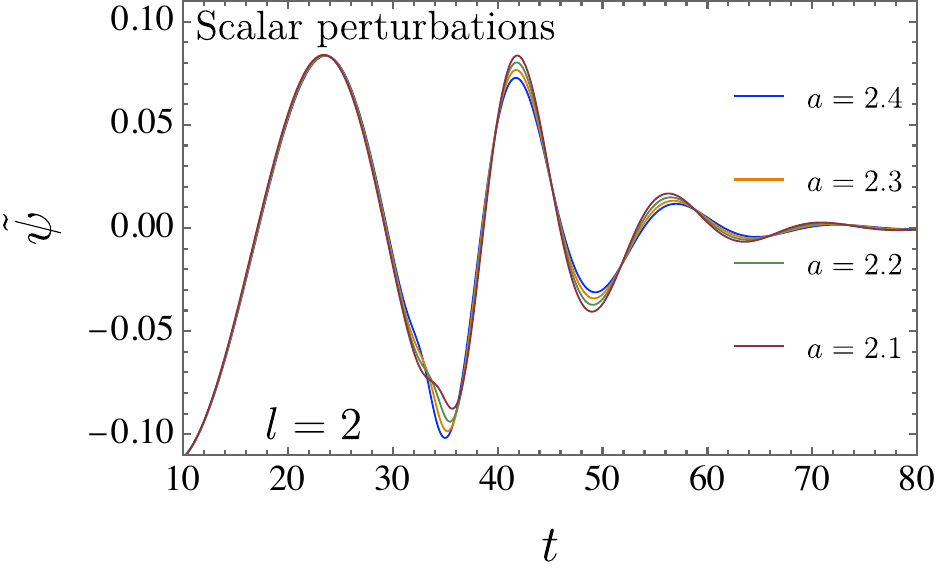}
    \caption{Numerical profiles of the scalar perturbation $\tilde{\psi}$ are shown for a {WH} with mass fixed at $M=1$, comparing several values of the parameter $a$, namely $2.1$, $2.2$, $2.3$, and $2.4$. The multipolar content is organized by panels corresponding to $l=1$ (left), and $l=2$ (right). All curves are obtained for a common choice $\Bar{\alpha}=0.7$.}
    \label{timedomainscalar0}
\end{figure}

The resulting time--domain signals are displayed in Fig.~\ref{timedomainscalar0} for $M=1$ and several values of the {throat} parameter, namely $a=2.1$, $2.2$, $2.3$, and $2.4$. Separate panels correspond to the angular multipoles $l=1$ (left), and $l=2$ (right). In all cases, the waveforms show a sequence of damped oscillations whose amplitudes decrease exponentially, which is characteristic of quasinormal ringing in BH backgrounds. A clear tendency emerges: for fixed $\Bar{\alpha}$, increasing $a$ leads to a slower attenuation of the oscillations, so that the perturbations persist over longer times. This behavior is consistent with the trends inferred from the quasinormal spectrum. {In addition, it is worth mentioning that the time-domain profiles are shown in units of $t/M$, with $M=1$. Also, the sharper reshaping visible in the $l=2$ panel around $t\approx 34$ might correspond to the transition between the prompt signal and the onset of the quasinormal ringing.}

\begin{figure}
    \centering
    \includegraphics[scale=0.51]{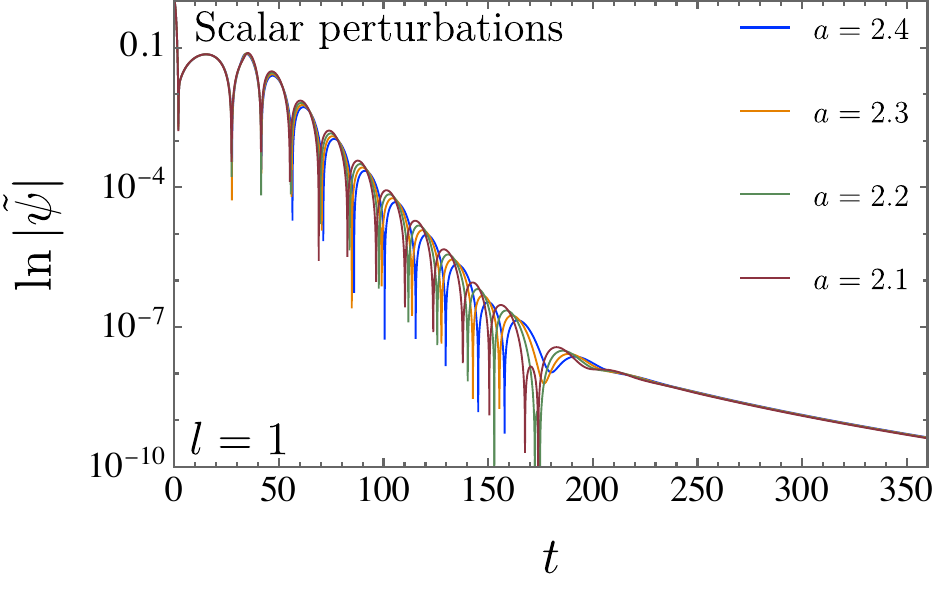}
     \includegraphics[scale=0.51]{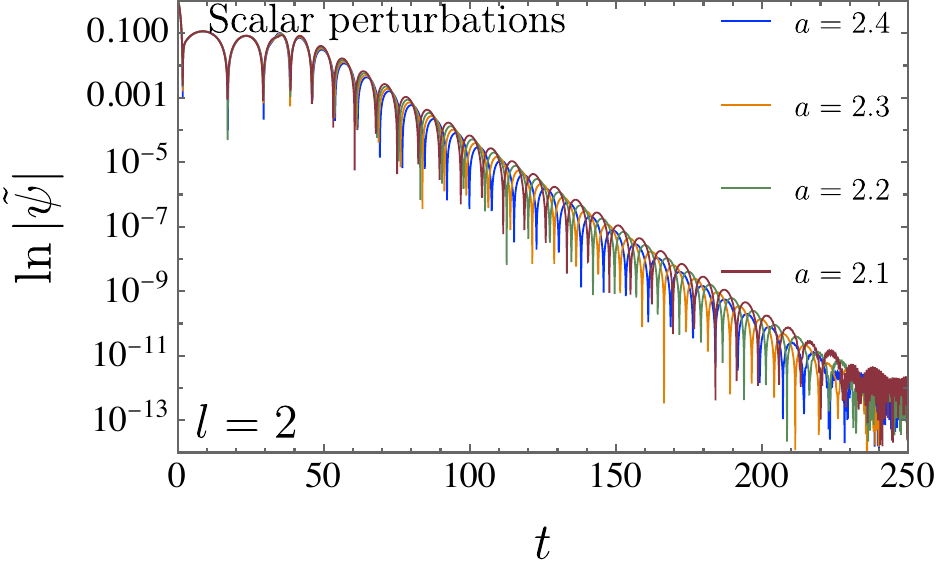}
    \caption{Logarithmic time--domain signals, $\ln|\tilde{\psi}|$, are presented for scalar field perturbations of a {WH} with mass set to $M=1$, for several values of the parameter $a$ ($2.1$, $2.2$, $2.3$, and $2.4$). The figure is arranged by multipole order, with $l=1$ in the left panel, and $l=2$ in the right panel. All results are obtained for a fixed choice $\Bar{\alpha}=0.7$.}
    \label{timedomainscalar1}
\end{figure}

The decay pattern is further clarified in Fig.~\ref{timedomainscalar1}, where the same evolutions are represented through $\ln|\tilde{\psi}|$ for identical choices of $a$ and $l$ (again at fixed $\Bar{\alpha}$). In this logarithmic scale, the quasinormal phase appears as nearly linear segments, signaling exponential damping, followed by a clear departure toward a different regime. The comparison between curves confirms that larger values of $a$ are associated with a milder decay, in agreement with the frequency--domain analysis presented earlier.

\begin{figure}
    \centering
    \includegraphics[scale=0.51]{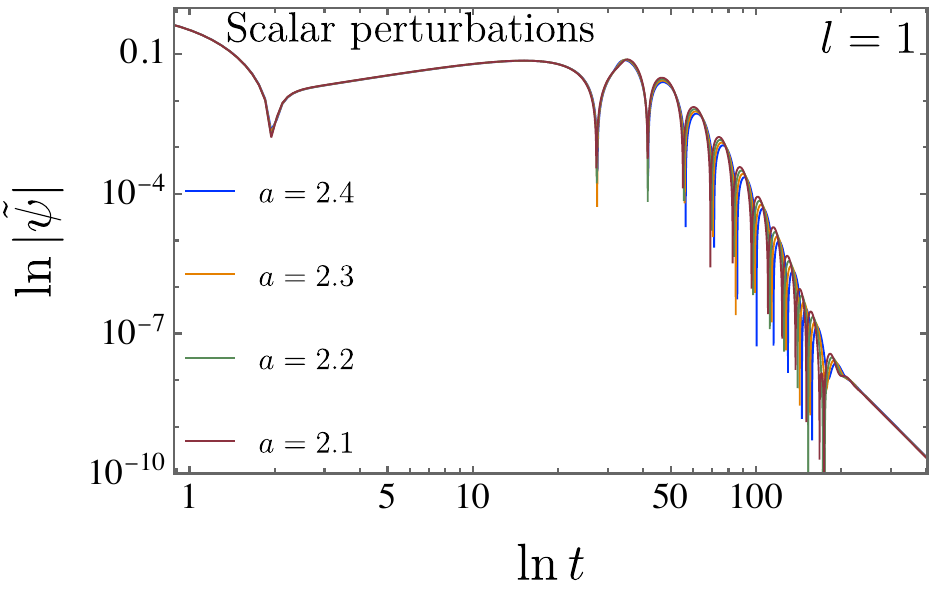}
     \includegraphics[scale=0.51]{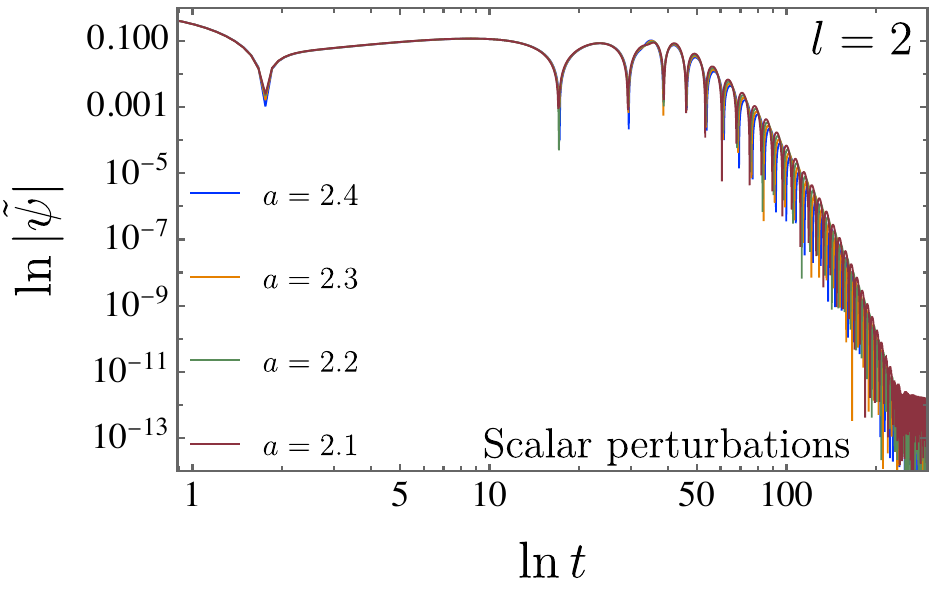}
    \caption{Asymptotic time--domain behavior of the scalar field $\tilde{\psi}$ is displayed using a double--logarithmic representation, plotting $\ln|\tilde{\psi}|$ versus $\ln t$ for a {WH} with mass fixed at $M=1$ and parameter values $a=2.1$, $2.2$, $2.3$, and $2.4$. The panels are arranged according to the angular multipole index, with $l=1$ in the left, and $l=2$ in the right panel. All curves are computed for the same choice $\Bar{\alpha}=0.7$.}
    \label{timedomainscalar2}
\end{figure}

Finally, Fig.~\ref{timedomainscalar2} highlights the evolution by plotting the field on a double--logarithmic scale, using the same panel arrangement for direct comparison. This representation makes explicit the crossover from the exponentially damped ringing to a slower, power--law falloff at late times. The presence of this tail behavior aligns with the standard expectations for scalar perturbations in BH spacetimes and completes the picture of the full temporal evolution therefore \cite{Molina:2010fb,Molina:2016tkr,konoplya2016wormholes,churilova2020arbitrarily,dutta2020revisiting}.

\section{Conclusion}\label{sec6}

In this work, we theoretically investigate the gravitational lensing effects on the spacetime of a traversable WH formed not because of the presence of an exotic matter source but rather through a geometric defect. Furthermore, the constant of a global monopole was incorporated, characterizing that this WH is embedded within this geometry. With regard to the {weak}-field regime, we show that the light deflection angle Eq. (\ref{13}) does not depend on the parameter that characterizes the throat of the WH, that is, it depends only on contributions from the global monopole. Otherwise, it would be equivalent to the angular deviation for the Schwarzschild BH. Still in the first part of the work, we analyzed the deflection of light in the strong-field regime, using the Bozza and Tsukamoto methodology to analyze mainly the part of the integration that presents logarithmic divergence and then control this divergence. In this way, we constructed an analytical expression for the divergent part of the integration of the angular deviation Eq. (\ref{23}) and, due to technical difficulties, we performed the numerical treatment of the regular part of the solution Eq. (\ref{24}), as well as for the total angular deviation, which can be observed in Figs. \ref{FORTE1} and \ref{FORTE2}.

In the second part of the work, we used the calculation of the angular deflection of light to construct the gravitational lensing equations, with the intention of correlating the observables to the relativistic images, both in the strong-field and weak-field regimes. For the strong-field regime, {we modelled the angular deflection using data for the BH at the centre of our Galaxy \cite{B12}}, and then created a data table \ref{TAB1} to understand how the charge on the global monopole is modifying the angular separation and the expression that relates the first-image flux to the others, in this case referring to the Schwarzschild BH. We also illustrate the behavior of these quantities through Fig. \ref{OBSERVAVEIS1}. In the same way, with regard to the weak-field regime, we consider data related to a bulge star \cite{BOJO} and analyze the effects of the correction due to the global monopole charge on the calculation of the Einstein ring and its angular position for some possible values of the monopole charge that can be observed in table \ref{TAB2}. Naturally, the values of these observables differ from those predicted by the Schwarzschild solution due to effects stemming from the global monopole charge, and we hope that, due to the analytical and numerical results developed throughout this work, we can contribute to the verification of new WH models that do not necessarily require an exotic matter source for their existence. Although the range of WH models already present in the literature that require exotic matter for their existence offer richer structures for analysis, in the case explored in this work, note that it does not even contribute to the calculation of the angular deviation in the weak-field regime, resulting in a simplified structure.

Regarding the dynamical response of the spacetime, the analysis of quasinormal modes and time–domain profiles showed that the deformation parameters, in particular the geometric parameter $a$ associated with the throat structure, directly influence the perturbative dynamics. As $a$ increased, the quasinormal spectrum exhibited a minor decrease in the real part of the frequencies together with a reduction in the magnitude of the imaginary part, indicating faster oscillations accompanied by weaker damping (for $\omega_{0}$ and $\omega_{1}$). This behavior was consistently reflected in the time--domain evolution, where larger values of $a$ led to a longer--lasting ringdown phase before the signal transitioned to the late--time tail governed by a power--law decay. Despite these quantitative modifications, the waveforms remained stable for all explored configurations.  {To make the role of the spacetime parameters more transparent, we summarize in Tab.~\ref{tab:paramdependence} which physical quantities depend explicitly on $M$, $a$, and $\bar{\alpha}$. }

\begin{table}[!h]
\begin{center}
\caption{\label{tab:paramdependence}
Summary of the dependence of the main physical quantities studied in this work on the spacetime parameters $M$, $a$, and $\bar{\alpha}$. Here, ``Yes'' means explicit dependence, while ``No'' means no explicit dependence in the corresponding analytical expression derived in the manuscript.}
\renewcommand{\arraystretch}{1.15}
\begin{tabular}{lccc}
\hline
\hline
Physical quantity & $M$ & $a$ & $\bar{\alpha}$ \\
\hline
Photon sphere radius $r_m$ & Yes & Yes & No \\
Shadow radius $R_{\mathrm{sh}}$ & Yes & No & No \\
Critical impact parameter $\beta_c$ / limiting angular position $\theta_{\infty}$ & Yes & No & No \\
Weak-field deflection angle $\delta\phi_{\mathrm{weak}}$ & Yes & No & Yes \\
Strong-field deflection angle $\delta\phi_{\mathrm{strong}}$ & Yes & Yes & Yes \\
Angular separation $s$ & Yes & Yes & Yes \\
Relative flux observable $\tilde r$  & No & No & Yes \\
Einstein ring angular position $\theta_E$ & Yes & No & Yes \\
Einstein ring radius $R_E$ & Yes & No & Yes \\
Scalar effective potential $V_s$ & Yes & Yes & Yes \\
Quasinormal frequencies $\omega_n$ & Yes & Yes & Yes \\
Time-domain profile $\tilde{\psi}(t)$ & Yes & Yes & Yes \\
\hline
\hline
\end{tabular}
\end{center}
\end{table}

As a further perspective, it appears worthwhile to examine how this WH configuration influences neutrino oscillations, in light of recent results reported in the literature \cite{Shi:2025tvu,Shi:2025plr,Shi:2025rfq,AraujoFilho:2025rzh}. Additionally, the accretion of matter onto the WH constitutes another relevant aspect that deserves detailed analysis. These and related directions are currently under active investigation.


\section*{Acknowledgments}
\hspace{0.5cm}

A. A. Araújo Filho is supported by Conselho Nacional de Desenvolvimento Cient\'{\i}fico e Tecnol\'{o}gico (CNPq) and Fundação de Apoio à Pesquisa do Estado da Paraíba (FAPESQ), project numbers 150223/2025-0 and 1951/2025. M. V. de S. Silva is supported by CNPq/PDE 200218/2025-5. R. L. L. Vit\'oria is supported by Conselho Nacional de Desenvolvimento Cient\'{\i}fico e Tecnol\'{o}gico (CNPq) and by Universidade Estadual do Marnh\~ao (UEMA), projects numbers 150420/2025-0 and by the EDITAL N. 102/2025-PPG/CPG/UEMA, respectively.

\section*{Conﬂict of Interest}
The author declares no conﬂict of interest.

\section*{Data Availability Statement}
The results are obtained through purely theoretical calculations and can beveriﬁed analytically; thus, this manuscript does not have associated data, or the data will not be deposited

	\bibliography{main}
\end{document}